\documentclass[aip,amsmath,amssymb,reprint]{revtex4-1}

\pdfoutput=1

\usepackage{dcolumn}
\usepackage{graphicx}
\usepackage[bottom]{footmisc}
\usepackage[includefoot]{geometry}

\begin{document}

\newcommand{\ud}{\mathrm{d}}
\newcommand{\de}[2]{\frac{\ud #1}{\ud #2}}
\newcommand{\diag}{\mathrm{diag}}
\newcommand{\bra}[1]{\langle #1|}
\newcommand{\ket}[1]{|#1\rangle}
\newcommand{\braket}[2]{\langle#1|#2\rangle}
\newcommand{\sech}{\mathrm{sech}}
\newcommand{\sgn}[1]{\mathrm{sgn} #1}
\newcommand{\pdern}[3]{\frac{\partial^#3#1}{\partial#2^#3}}
\newcommand{\pder}[2]{\frac{\partial#1}{\partial#2}}
\let\Eps\varepsilon
\newcommand{\Deltalpha} {\Delta\!\alpha}


\title{Mean flow modelling in high-order nonlinear Schr\"odinger equations} 



\author{Alexis Gomel}
\altaffiliation[Presently at~]{Department of Mathematics and Statistics, University of Reading, Reading, UK}
\affiliation{Group of Applied Physics and Institute for Environmental Sciences, University of Geneva, 66 Bd Carl-Vogt, CH-1211 Geneva 4, Switzerland}
\author{Corentin Montessuit}
\affiliation{Group of Applied Physics and Institute for Environmental Sciences, University of Geneva, 66 Bd Carl-Vogt, CH-1211 Geneva 4, Switzerland}
\author{Andrea Armaroli} 
\altaffiliation[Presently at ]{Department of Engineering, University of Ferrara, via Saragat 1, I-44122 Ferrara, Italy}
\affiliation{Group of Applied Physics and Institute for Environmental Sciences, University of Geneva, 66 Bd Carl-Vogt, CH-1211 Geneva 4, Switzerland}
\author{Debbie Eeltink}
\affiliation{Department of Mechanical Engineering, Massachusetts Institute of Technology, Cambridge, Massachusetts 02139, USA}
\affiliation{Department of Engineering Science, University of Oxford, Parks Road, Oxford OX1 3PJ, UK}
\affiliation{Laboratory of Theoretical Physics of Nanosystems, EPFL, CH-1015 Lausanne, Switzerland }
\author{Amin Chabchoub}
\affiliation{Disaster Prevention Research Institute \& Hakubi Center for Advanced Research, Kyoto University, Kyoto, 606-8501 Kyoto, Japan}
\affiliation{School of Civil Engineering, The University of Sydney, Sydney, NSW 2006, Australia}
\author{J\'er\^ome Kasparian}
\author{Maura Brunetti}
\email[Corresponding author: ]{maura.brunetti@unige.ch}
\affiliation{Group of Applied Physics and Institute for Environmental Sciences, University of Geneva, 66 Bd Carl-Vogt, CH-1211 Geneva 4, Switzerland}


\date{\today}

\begin{abstract}
The evaluation and consideration of the mean flow in wave evolution equations are necessary for the accurate prediction of fluid particle trajectories under wave groups, with relevant implications in several domains, from the transport  of pollutants in the ocean, to the estimation of energy and momentum exchanges between the waves at small scales and the ocean circulation at large scale. 
We derive an expression of the mean flow at finite water depth which, in contrast to other approximations in the literature,  accurately accords with the deep-water limit at third order in steepness, and is equivalent to second-order formulations in intermediate water. We also provide envelope evolution equations at fourth order in steepness for the propagation of unidirectional wave groups either in time or space that include the respective mean flow term. The latter, in particular, is required for accurately modelling experiments in water wave flumes in arbitrary depths. 
\end{abstract}

\pacs{}

\maketitle 

\section{Introduction}
\label{sec:1}

As waves evolve on the ocean surface, they induce a mean flow to the fluid particles, particularly when nonlinear effects are taken into account.
Near the surface, fluid particles experience a net horizontal displacement in the same direction as the water wave propagation, the so-called Stokes drift~\cite{stokes1847,vandenBremer2018}, that decays with depth. A return  flow in the fluid column along vertical and horizontal directions guarantees the conservation of the water-mass transport, causing localised variations in the mean water level under wave groups and the associated propagation of infragravity waves~\cite{Kalish2021}  
at finite water depth. 
An accurate description of the mean flow is thus necessary for the proper reconstruction of the fluid particle trajectories underneath water waves~\cite{Calvert2019,carter2020}. Since the wave-induced mean flow is associated to the transport of energy, momentum and other tracers~\cite{dibenedetto2020}, such as pollutants like plastic~\cite{cozar2014} or offshore oil spill~\cite{christensen2018}, it is relevant for environmental studies. Moreover, it has an impact on the general circulation of the ocean at large scales 
and its modelling~\cite{babanin2006,Onink2019,vanSebille2020,Cunnigham2022}, and on the nearshore circulation~\cite{LONGUETHIGGINS1964,Kalish2021}, in particular in shoaling regions~\cite{janssenSloping2003,battjes2004}, while it is affected by the presence of background shear currents~\cite{monismith2007,curtis2018}.

Since Stokes drift and return flow are two phenomena occurring at second-order in steepness, they are taken into account in the finite water depth nonlinear Schr\"odinger equation (NLS), which describes the evolution of the envelope of narrow-banded wave packets and can be obtained using a multi-scale development of water surface elevation and velocity potential at third-order in steepness~\cite{BenneyRoskes1969,HasimotoOno1972,DaveyStewartson1974,doi:10.1142/5566}, or taking the narrow-banded limit of the Zakharov equation using an Hamiltonian approach~\cite{Zakharov1968,janssen2007intermediate}.  
Indeed, at third-order in steepness, it was shown~\cite{DaveyStewartson1974} that the mean flow term comes into play as a modification of the nonlinear coefficient, giving rise to a transition from focusing to defocusing regimes at the critical value $k_0 h\approx 1.363$, where $k_0$ is the carrier wavenumber and $h$ the depth. However, this contribution to the nonlinear coefficient disappears in the deep-water limit. At fourth-order in steepness, the accurate formulation of the mean flow term needs to be accounted for, in both, finite and infinite depth waters~\cite{Dysthe1979}.

Fourth-order terms in the NLS equation are necessary to explain features like asymmetrical evolution of spectra~\cite{PhysRevLett.111.054104}, and asymmetries in the waveform~\cite{trulsen2001,shemer2010,GoulletChoi2011,zhang2014,Armaroli2017}, especially when inevitable wave focusing is at play~\cite{slunyaev2013}. 
Several versions of high-order wave envelope evolution equations exist in the literature, that can be obtained when applying the multiple scales development using steepness and bandwidth with different orders of magnitude~\cite{Dysthe1979,Janssen1983,brinch1986,Trulsen1996}, or the Hamiltonian approach~\cite{Stiassnie1984,DesbarmaDas2005,janssen2007intermediate,gramstad_trulsen_2011,gramstad_2014}.
The main difference between all these equations is the treatment and approximation of the mean flow term.  

Here, we will focus on narrow-banded unidirectional wave packets where steepness and bandwidth can be considered as parameters of the same order, like in the finite-depth developments described in Refs.~\onlinecite{Sedletsky2003} (denoted as Sed03 in the following) and \onlinecite{Slunyaev2005} (Slu05). The  challenge of such developments is the fact that they do not reduce to the Dysthe equation (Ref.~\onlinecite{Dysthe1979}, Dys79) in the deep-water limit, which has been shown to well reproduce wave tank experiments~\cite{GoulletChoi2011}. 
We will show that this convergence depends on how the mean flow is approximated. 
Moreover, we will provide finite-depth envelope equations at fourth order in steepness in both, space-like and time-like formulations, with correct limiting expressions in deep water. 
In particular, such unidirectional time-like equations that allow a continuous scaling from intermediate to the deep water limit are relevant for reproducing wave tank experiments with high accuracy in arbitrary depths. In contrast, directional sea states are typically found in the open ocean,  and this restricts the applicability of the the above unidirectional modelling equations.

The paper is organised as follows. In Sec.~\ref{spacelike}, we will derive the expression of the mean flow to be inserted in the envelope equation at fourth order in steepness for the evolution in time. We will compare this expression with the ones already existing in the literature, and show that it indeed correctly describes the mean flow in the whole range of water depths, {\it i.e.}, from intermediate to deep water regimes. In Sec.~\ref{timelike}, we will provide an envelope equation for the water wave evolution in space, relevant for modelling for instance water tank experiments, and the corresponding 
mean flow term, at fourth order in steepness. Again, we will perform the comparison for various water depth scenarios. Finally, in Sec.~\ref{conclusion} we will summarise our findings.   
\section {Fourth-order equation: propagation in time}
\label{spacelike}

The multi-scale approach has been used to derive at fourth order in steepness the following equation, that describes the evolution of the envelope $U$ of a uni-directional progressive gravity wave packet propagating in time 
on the free surface of a homogeneous liquid with depth $h$~\cite{Sedletsky2003, Slunyaev2005,gandzha2014} 
\begin{widetext}
\begin{equation}
		i\left(\pder{U}{t} + c_\mathrm{g} \pder{U}{X}\right) 
		+ \hat\alpha\pdern{U}{X}{2}  \underbrace{-\hat\beta|U|^2U}_\text{incl.~Mean Flow}
		 =i\biggl( {\hat\alpha_3 \pdern{U}{X}{3}}
			  \underbrace{- \hat\beta_{21} |U|^2 \pder{U}{X}
			   {-  \hat\beta_{22} U^2 \pder{U^*}{X}}}_\text{incl.~Mean Flow}\biggr).
			\label{eq:HONLSspace}
\end{equation}
\end{widetext}
The explicit formulation of the group velocity $c_\mathrm{g}$, and of all dispersive and nonlinear coefficients $\hat\alpha$, $\hat\beta$, $\hat \alpha_3$,  $\hat\beta_{21}$, and $\hat\beta_{22}$ is given in App.~\ref{app:A} and \ref{app:B}. The surface tension has been neglected and the fluid is considered as irrotational. 
The sign convention follows Sed03: the surface elevation is reconstructed at leading order from the envelope using\footnote{Note that Slu05 uses the opposite convention, giving rise to different signs in some terms with respect to Sed03.} 
 $\eta(X,t) = \frac{1}{2} [U(X,t) \exp(i (k_0 X- \omega_0 t)) + {\rm{c.~c.}}]$, where $\omega_0$ and $k_0$ are the angular frequency and wavenumber of the carrier wave, respectively.  

\begin{figure}[ht!]
\centering
\includegraphics[width=0.49\textwidth]{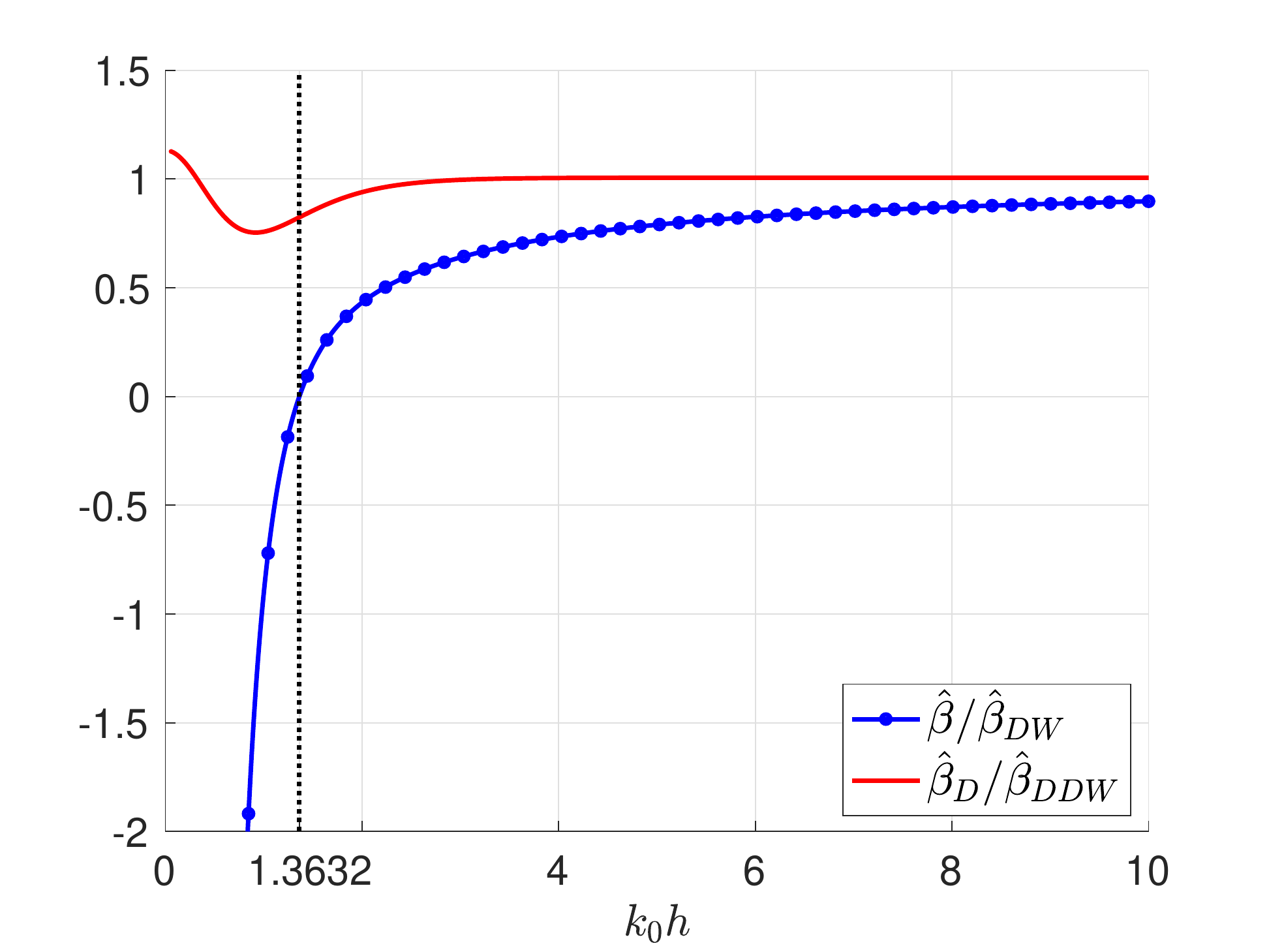}
\caption{Third-order nonlinear coefficients $\hat\beta$ and $\hat\beta_D$ in Eqs.~(\ref{eq:HONLSspace}) and (\ref{eq:HONLSspace2}), respectively, normalised with respect to their deep-water values, which are equal. Zero crossing for $\hat \beta$ is at $k_0h = 1.363$, shown by the dotted vertical line.}
\label{fig:betaD}
\end{figure}

The previous Eq.~(\ref{eq:HONLSspace}) is referred to as the `space-like equation' since dispersion is in space. It can also be written in the following equivalent form~\cite{Sedletsky2003} in a reference frame moving with the group velocity $x = X -c_g t$:
\begin{widetext}
\begin{equation}
i \pder{U}{t}  + \Eps\left(\hat\alpha\pdern{U}{x}{2} - \hat\beta_D|U|^2U\right) =i\Eps^2\left({\hat\alpha_3 \pdern{U}{x}{3}}
{- \omega_0 k_0 \tilde Q_{41} |U|^2 \pder{U}{x}}
{- \omega_0 k_0 \tilde Q_{42} U^2 \pder{U^*}{x}}\right) + \underbrace{\frac{\mu_g k_0}{4\sigma}U\pder{\phi_0}{x} }_{\text{Mean Flow}}
\label{eq:HONLSspace2}
\end{equation}
\end{widetext}
where $\sigma = \tanh(k_0h)$, $\mu_g$ is given in Eq.~(\ref{eq:mug}), and the high-order nonlinear coefficients $\tilde Q_{41}$ and $\tilde Q_{42}$ in Eqs.~(\ref{eq:tildeQ41}) and (\ref{eq:tildeQ42}), respectively. The nonlinear term $\hat\beta_D$ is a positive function (see Fig.~\ref{fig:betaD}), given in Eq.~(\ref{eq:hatbetaD}), and since the dispersion coefficient $\hat \alpha$ is negative, it seems that the characterisation of the focusing and defocusing regime is somehow lost in the present formulation.

Notice that we explicitly introduce the scaling parameter $\Eps$, a dummy variable which is set to 1 at the end, that is helpful for grouping terms of the same order in steepness. The last term, depending on the zero harmonic of the velocity potential $\phi_0$, is the mean flow term that, from the multi-scale development, takes the following form (see Eqs.~(35) and (57) in Sed03): 
\begin{eqnarray}
 \pder{\phi_0}{x} &=&   \Eps \frac{\omega_0}{2} \frac{\mu_g k_0}{\sigma \nu}  |U|^2 \nonumber \\ 
 &-& i \Eps^2 \frac{4\omega_0 \sigma}{\nu} \tilde q_{40S} \left(U \pder{U^*}{x}- U^* \pder{U}{x}\right) \hspace{1cm}
\label{eq:mean4order}
\end{eqnarray}
with $\nu$ given in Eq.~(\ref{eq:nu}) and $\tilde q_{40S}$ in Eq.~(\ref{eq:q40S}).  Its contribution can be included into the coefficients $\hat\beta_D$, $\tilde Q_{41}$, $\tilde Q_{42}$ to obtain $\hat\beta$, $\hat\beta_{21}$, $\hat\beta_{22}$, respectively,  in Eq.~(\ref{eq:HONLSspace}) 
(see steps from Eq.~(40) to (42) and from Eq.~(64) to (66) in Sed03). Thus, the 
mean flow term is already taken into account in Eq.~(\ref{eq:HONLSspace}) at fourth order in steepness through the nonlinear coefficients.  

We remark that from Eq.~(\ref{eq:mean4order}), in the deep-water limit, $\partial\phi_{0}/\partial x \to 0$ since $\nu \to -\infty$ while all the other coefficients are finite. This agrees with the developments at third order in steepness~\cite{DaveyStewartson1974,Peregrine1983}. 
Note that in the 1D shallow-water case the mean flow term modifies the nonlinear term, see Eq.~(2.21) in Ref.~\onlinecite{DaveyStewartson1974}, while its contribution vanishes in the deep-water limit. However, the fact that the contribution of mean flow remains zero in deep water at higher-order approximation in steepness is not ideal, as we will discuss in the following subsection.  

\subsection{Mean flow in the deep-water limit} 
\label{sec:HilbertTerm}

In the Dysthe equation (Dys79), {\it i.e.}, the evolution equation obtained in the deep-water limit with the multi-scale method at fourth order in steepness, an additional term corresponding to the wave-induced mean flow usually appears in the literature in both, space-like~\cite{Dysthe1979,Janssen1983,LoMei1985} and time-like 1D formulations~\cite{TrulsenDysthe1997,kitShemer2002,OnoratoOsborneEtAl2005,GoulletChoi2011,Eeltink2017}. Let us consider for the moment the propagation in time. The Dysthe equation reads:
\begin{widetext}
\begin{equation}
i\pder{U}{t} - \frac{\omega_0}{8k_0^2}\pdern{U}{x}{2} - \frac{\omega_0 k_0^2}{2} |U|^2U = i\frac{\omega_0}{16k_0^3} \pdern{U}{x}{3} -
 i \frac{\omega_0 k_0}{2}  \biggl(3 |U|^2 \pder{U}{x}
+  \frac{1}{2} U^2 \pder{U^*}{x} + \underbrace{\frac{2i}{\omega_0} U \pder{\phi_{0}}{x}}_\text{Mean Flow}\biggr)
\label{eq:meanflow2SPACELIKE}
\end{equation}
\end{widetext}
In this deep-water limit, the mean-flow term  is written as~\cite{OnoratoOsborneEtAl2005,GoulletChoi2011}: 
\begin{equation}
\pder{\phi_0}{x}|_{z=0} =   
\frac{\omega_0}{2} {\mathcal{H}_x} \left[\pder{|U|^2}{x}\right], \hspace{1cm} 
\label{HilbertTerm}
\end{equation} 
where $\mathcal{F}_x$ is the Fourier transform in space, and $\mathcal{H}_x$ the Hilbert transform,  
$\mathcal{H}_x[\eta]= \mathcal{F}^{-1}_x\left[+i~\sgn(k)\mathcal{F}_x\left[\eta\right]\right]$, 
$\eta$ being a function of $x$. The role of the mean flow Hilbert term in the evolution of pulsating wave packets~\cite{PhysRevLett.111.054104} and narrow-banded irregular waves~\cite{shemer2010,GoulletChoi2011} has been shown in several experiments in deep water,   
and its presence is required for a correct modelling~\cite{trulsen2001}.  
 
Janssen~\cite{Janssen1983} suggests that the system can be closed by solving the equations for $\phi_0$ as a function of the envelope $U$ in Fourier space. It is instructive to report here the explicit derivation of the  Hilbert term since we will use analogous developments in the following.
The zero harmonic of the velocity potential $\phi_0$ satisfies the Laplace equation in the entire water column with boundary conditions at the surface and at the bottom, giving a set of equations that  constitutes the following Neumann problem:  
\begin{subequations}
\begin{align}
   \pdern{\phi_0}{z}{2} + \pdern{\phi_0}{x}{2}  &= 0 &- \infty < z \leq 0  \\
   \pder{\phi_0}{z} &=  \frac{\omega_0}{2}\pder{|U|^2}{x} & z=0 \label{eq:surface} \\
   \pder{\phi_0}{z} &= 0 & z \rightarrow -\infty \label{eq:bottom}
\end{align}
\end{subequations}
Substituting the Fourier transform of the mean flow: 
\begin{eqnarray}
&&\mathcal{F}_x \phi_0(x,z,t)  = \hat\phi_0(k,z,t) \nonumber \\ 
&=& \frac{1}{\sqrt{2\pi}}\int_{-\infty}^{\infty} \phi_0(x,z,t)e^{ikx} dx    
\end{eqnarray}
into the Laplace equation gives $\partial^2 \hat \phi_0/\partial z^2 = k^2 \hat\phi_0$ whose solution is 
\begin{equation}
\hat \phi_0(k,z,t) = C_1 e^{|k|z}   
\end{equation}
with $C_1$ independent of $z$. This satisfies the bottom boundary condition for $z\to -\infty$. 

Inserting this expression in Eq.~(\ref{eq:surface}) gives: 
\begin{equation}
|k| C_1 e^{|k| z}|_{z=0} = |k| C_1= \frac{\omega_0}{2} \mathcal{F}_x \left[\pder{|U|^2}{x}\right], 
\end{equation}
from which $C_1$ is obtained, and thus: 
\begin{equation}
\hat \phi_0(k,z,t) = \frac{1}{|k|}\frac{\omega_0}{2} e^{|k|z} \mathcal{F}_x\left[\pder{|U|^2}{x}\right]\,. 
\end{equation}
Now the derivative with respect to $x$ at $z=0$ 
in Fourier space is given by: 
\begin{eqnarray}
&&\mathcal{F}_x\left[\pder{\phi_0}{x}\right]|_{z=0} = ik \hat\phi_0|_{z=0} \nonumber \\ 
&=&
i\frac{\omega_0}{2}\sgn(k) \mathcal{F}_x \left[\pder{|U|^2}{x}\right] 
\end{eqnarray} 
and finally, moving back to the direct physical space, the Hilbert term of Eq.~(\ref{HilbertTerm}) is recovered. 

\subsection{Multi-scale development and Hilbert term}

As discussed, the multi-scale approach of Sed03 gives Eq.~({\ref{eq:mean4order})}, which in the deep-water limit reduces to 
\begin{equation}
\frac{\partial }{\partial x}(\phi_{01}+\phi_{02}) = 0 \, . 
\label{eq:multiscale}
\end{equation}
On the other hand, the mean flow can be written as the Hilbert term of Eq.~(\ref{HilbertTerm}) in the Dysthe equation~\cite{Dysthe1979,Janssen1983}. Thus, there is the need to reconcile these results. This can be done as follows.
In the derivation of the Hilbert term, we use the Laplace equation for $\phi_0$ that is the {\it complete} mean flow, and not just its approximation at second order in steepness as in Eq.~(\ref{eq:multiscale}).
Indeed, the Laplace equation at third order for the mean flow is: 
\begin{equation}
\frac{\partial^2 \phi_{03}}{\partial z^2} +\frac{\partial^2\phi_{01}}{\partial x^2} = 0  .  
\end{equation}
When integrated in $z$, this gives
\begin{equation}
\frac{\partial \phi_{03}}{\partial z} = -(z+h) \frac{\partial^2\phi_{01}}{\partial x^2},
\label{eq:phi03z}
\end{equation}
using the fact that 
$\partial \phi_{01}/\partial z = 0$ and imposing the bottom boundary condition.
From the multi-scale development, the following expression can be obtained at third-order in steepness (see Eqs.~(2.12) and (2.14) in~Ref.~\onlinecite{DaveyStewartson1974}):
\begin{eqnarray}
\frac{c_g^2}{g}\frac{\partial^2\phi_{01}}{\partial x^2} +\frac{\partial\phi_{03}}{\partial z}  &=& \frac{\omega_0}{2\sigma} (1+C_{FD}) \pder{|U|^2}{x} \nonumber \\ 
&=&  D' \pder{|U|^2}{x}, 
\label{eq:calvert}
\end{eqnarray}
with the coefficient $C_{FD}$ defined in Eq.~(\ref{eq:CFD}) and in Ref.~\onlinecite{Calvert2019}, and 
where $D'=(1+C_{FD})\,\omega_0/(2\sigma)=\mu_g\omega_0/(8\sigma^2)$ (see definitions in App.~\ref{app:A}). 
Using Eq.~(\ref{eq:phi03z}) at $z=0$,  Eq.~\eqref{eq:calvert} reduces to
\begin{equation}
\frac{\partial^2\phi_{01}}{\partial x^2} = -\frac{D}{h} \pder{|U|^2}{x}, 
\label{eq:calvertRed}
\end{equation}
which corresponds to Eqs.~(33)-(34) in Sed03 and where $D$ is defined in Eq.~(\ref{eq:Ddef}).
Eq.~(\ref{eq:calvertRed}) can also be written, using Eq.~(\ref{eq:phi03z}), as:
\begin{equation}
\frac{\partial\phi_{03}}{\partial z} = D\pder{|U|^2}{x}. 
\label{eq:surfbc}
\end{equation}
Taking the deep-water limit gives Eq.~(\ref{eq:surface}), 
\textit{i.e.}, the boundary condition at the surface in the Neumann problem. Since $\partial \phi_{01}/\partial z = 0$ and $\partial \phi_{02}/\partial z = 0$, such expression is valid at third order in steepness for the mean flow. 

Thus, in the derivation of the Hilbert term, the complete mean flow (in the Laplace equation) is considered together with the mean flow at third order in steepness in the surface boundary condition. Consequently, a `hybrid' relation as described by Eq.~(\ref{HilbertTerm}) is obtained, which is different from Eq.~(\ref{eq:multiscale}) where only the first terms in the development of the mean flow at the surface are taken into account. In other words, the expression in Eq.~(\ref{HilbertTerm}) is inherently nonlocal. The same occurs in the nonlinear terms of the super-compact model \cite{dyachenko2017}, from which the Dysthe equation can be derived. The terms in Sed03 and Slu05 are instead local, as they only involve the surface, and not the entire water column. This is obviously an approximation, since the mean flow does involve a body of fluid which is not immediately at the surface, as clearly shown in field measurements~\cite{Kalish2021}.  

\subsection{Mean flow term in arbitrary depth}
\label{subsec:arbitrary}

We now repeat the procedure used in Sec.~\ref{sec:HilbertTerm} for the case 
of intermediate water.
 We will consider two cases that differ based on the considered surface boundary condition:  the Neumann problem is solved using the condition given by Eq.~(\ref{eq:surfbc}) in {\bf Case 1}; using Eq.~(\ref{eq:calvert}) in {\bf Case 2}. Replacing the expression of $\partial \phi_0/\partial x$ that is obtained in each case into the last term of Eq.~(\ref{eq:HONLSspace2}) gives the final high-order NLS equation in arbitrary finite depth and space-like form.

\subsubsection*{Case 1}

Moving as before to the Fourier space, the Laplace equation is $\partial^2 \hat \phi_0/\partial z^2 = k^2 \hat\phi_0$ and  its solution is given by:
\begin{equation}
\hat \phi_0(k,z,t) = C_1 e^{|k|(z+h)}+C_2 e^{-|k|(z+h)}. 
\end{equation}
Imposing the boundary condition at the bottom, Eq.~(\ref{eq:bottom}), gives:
\begin{equation}
\pder{\hat\phi_0}{z}|_{z=-h} =|k|(C_1-C_2 )=0.
\end{equation}
Thus, we have $C_{1}=C_{2}=C/2$ and:
\begin{equation}
\hat \phi_0(k,z,t) = C\cosh\big(|k|(z+h)\big). 
\label{solutionIW}
\end{equation}
Inserting this in Eq.~(\ref{eq:surfbc}) gives:
\begin{equation}
    |k|C\sinh\big(|k|(z+h)\big)|_{z=0}= D \mathcal{F}_x\left[\pder{|U|^2}{x}\right], 
\end{equation}
from which one obtains $C$ and therefore:
\begin{equation}
    \hat\phi_{0}=\frac{1}{|k|}\frac{\cosh\big(|k|(z+h)\big)}{\sinh\big(|k|h\big)} D \mathcal{F}_x\left[\pder{|U|^2}{x}\right]. 
    \label{eq:LaplaceSol}
\end{equation}
At $z=0$, this gives:
\begin{equation}
\hat\phi_{0}|_{z=0}=D\frac{\coth\big(|k|h) }{|k|} \mathcal{F}_x\left[\pder{|U|^2}{x}\right]\,.
\label{eq:LaplaceSol2}
\end{equation}
Now, the derivative with respect to $x$ at $z=0$ is given by:
\begin{eqnarray}
&&\mathcal{F}_x\left[\pder{\phi_0}{x}\right]|_{z=0} = ik \hat\phi_0|_{z=0} \nonumber \\
&=& i D\frac{\sgn(k) }{\tanh(|k|h)}\mathcal{F}_x \left[\pder{|U|^2}{x}\right].
\label{eq:Fdx}
\end{eqnarray} 
Moving back to the direct physical space, we finally get:
\begin{equation}
    \pder{\phi_0}{x}=D\mathcal{F}^{-1}_x\Bigg\{\frac{i}{\tanh(k h)} \mathcal{F}_x \left[\pder{|U|^2}{x}\right]\Bigg\}. 
\label{eq:IWspace}    
\end{equation}

\subsubsection*{Case 2} 

The surface boundary condition is now Eq.~(\ref{eq:calvert}). Note that the relation $\phi_{03z}=-h\phi_{01xx}$ has not been used to simplify the l.h.s.~of this equation and both mean flow terms are thus considered being of the same order, see Eq.~(13) in Ref.~\onlinecite{Calvert2019}.
 
Inserting the generic form of the solution in intermediate water, \textit{i.e.}, Eq.~(\ref{solutionIW}), into the surface boundary condition, Eq.~\eqref{eq:calvert}, gives: 
\begin{widetext}
\begin{equation} 
	C\left.\left[	\frac{c_g^2}{g} (-k^2)\cosh(|k|(z+h)) +|k|\sinh(|k|(z+h))  \right]\right\rvert_{z=0} = D'\mathcal{F}_x \left[\pder{|U|^2}{x}\right],
\end{equation}
from which we obtain the following expression for $C(k,t)$:
\begin{equation}
	C(k,t) = \frac{D'}{k \tanh(kh)\left[1-
 c_g^2 k /(g\tanh(kh))\right]}
 \frac{1}{\cosh(kh)} \mathcal{F}_x\left[\pder{|U|^2}{x}\right],
\end{equation}
where we used $|k|\tanh (|k|h) = k \tanh (kh)$, and $c_g$ is the wave group speed at the carrier wavenumber. Performing analogous steps as in the previous case, 
the final expression for the Euler horizontal velocity in $z=0$ is:
\begin{equation}
\pder{\phi_0}{x}= D'\mathcal{F}^{-1}_x\left\{\frac{i}{\tanh{(kh)}\left[1-c_g^2k/(g\tanh(kh))\right]}\mathcal{F}_x\left[\pder{|U|^2}{x}\right] \right\} .
\label{eq:cal2}
\end{equation}
\end{widetext}
Note that this expression coincides with Eq.~(15) in Ref.~\onlinecite{Calvert2019} (apart from a sign that is a typo in that latter equation). \\

By performing the derivative in $x$ on the r.h.s.~of Eq.~(\ref{eq:IWspace}) or (\ref{eq:cal2}), considering that $D = D'/(1-c_g^2/(gh))$ and using $\coth y = 1/y  + O(y)$, we get in both {\bf Case 1} and {\bf 2} for small $kh$ numbers:
\begin{eqnarray}
    \pder{\phi_0}{x} &\sim& D \mathcal{F}^{-1}_x 
    \Bigg[i(ik)\coth(k h) \mathcal{F}_x[|U|^2]\Bigg] \nonumber \\
    &\sim& D \mathcal{F}^{-1}_x 
    \Bigg[-\frac{1}{h} \mathcal{F}_x[|U|^2]\Bigg] \nonumber \\
    &=& - \frac{D}{h} |U|^2=\frac{\omega_0}{2}\frac{\mu_g k_0}{\sigma \nu} |U|^2\, . 
\label{eq:smallk}    
\end{eqnarray}
Thus, we recover the first term on the r.h.s.~of Eq.~(\ref{eq:mean4order}). The NLS nonlinear coefficient is also recovered when the mean flow term is added to the third-order nonlinear term in Eq.~(\ref{eq:HONLSspace2}), since $\hat\beta =\hat\beta_D  
+  D^2 \nu/(2h^2\omega_0)$. Hence, the defocusing regime is recovered through the inclusion of the mean flow term. 

\subsection{Numerical comparisons}
\label{sec:comparisons}

We compare the expressions for the horizontal velocity $\partial \phi_0/\partial x$ listed in Table~\ref{tab:table1} with the sub-harmonic velocity potential $\phi_{20}$ at second-order in steepness and its horizontal derivative calculated using the Dalzell analytical method (see the original paper, Ref.~\onlinecite{dalzell1999} (Dal99), and the explicit formulae reported in the Appendix of Ref.~\onlinecite{LiLi2021}).

\begin{table}
\caption{\label{tab:table1} 
List of mean flow terms in space-like formulation: $\partial \phi_0/\partial x$. }
\begin{ruledtabular}
\begin{tabular}{lll}
{\bf Case 1} & Eq.~(\ref{eq:IWspace}) & 3rd-order, nonlocal \\
{\bf Case 2} & Eq.~(\ref{eq:cal2}) & 3rd-order, nonlocal \\
Dys79~\cite{Dysthe1979}  & Eq.~(\ref{HilbertTerm})  & 3rd-order, nonlocal \\
&& deep water \\
Sed03~\cite{Sedletsky2003}  & Eq.~(\ref{eq:mean4order})  & 3rd-order \\
Sed03  & Eq.~(\ref{eq:smallk})  & 2nd-order \\
Slu05~\cite{Slunyaev2005}  & Eq.~(\ref{eq:mean4orderSLUN})  & 3rd-order \\
\end{tabular}
\end{ruledtabular}
\end{table}

For benchmarking and validation purposes, we use the same parameters as in Ref.~\onlinecite{LiLi2021} (case C in their Table II), namely a Gaussian (amplitude) spectrum $S(k)$ with peak wavenumber 
$k_0 = 0.0277$~m$^{-1}$, wavelength $\lambda_0 = 2\pi/k_0$, standard deviation of the spectrum given by symmetrical values $k_w = k_{w1} = k_{w2} = 0.27 k_0$, steepness $\varepsilon = 0.3$ and different values of the normalised depth $k_0 h$ in the range $[0.5, 50]$, thus, the case of finite and infinite water depth conditions. 
The angular frequency is calculated from $\omega_0^2 = gk_0 \tanh(k_0h)$.
In particular, the surface elevation at first order in steepness given by the superposition of $N = 30$ waves is used to calculate the intensity of wave envelope, {\it i.e.} $|U|^2$, and then, the horizontal velocity $\partial \phi_0/\partial x$ through the different relations, as listed in Table~\ref{tab:table1}. We consider a focused wave group, composed of $N$ individual sinusoidal wave components that are in phase at a single point in time and space~\cite{orszaghova2014} (the origin in our case), and a random sea state by using a uniform distribution for the phases. An example of sea state realization is given in App.~\ref{app:D}.
   
\begin{figure*}[!]
\includegraphics[width=0.495\textwidth]{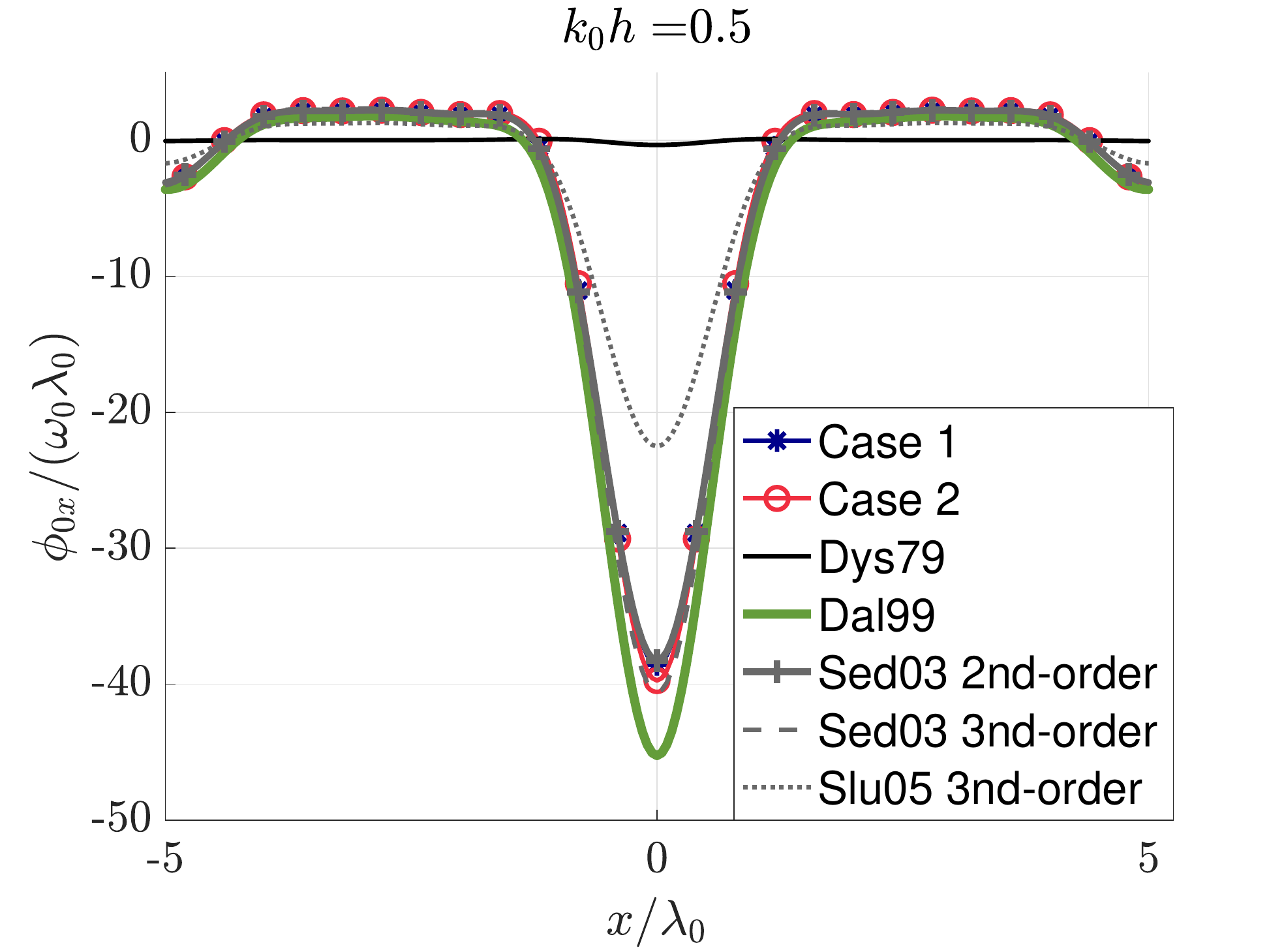}
\includegraphics[width=0.495\textwidth]{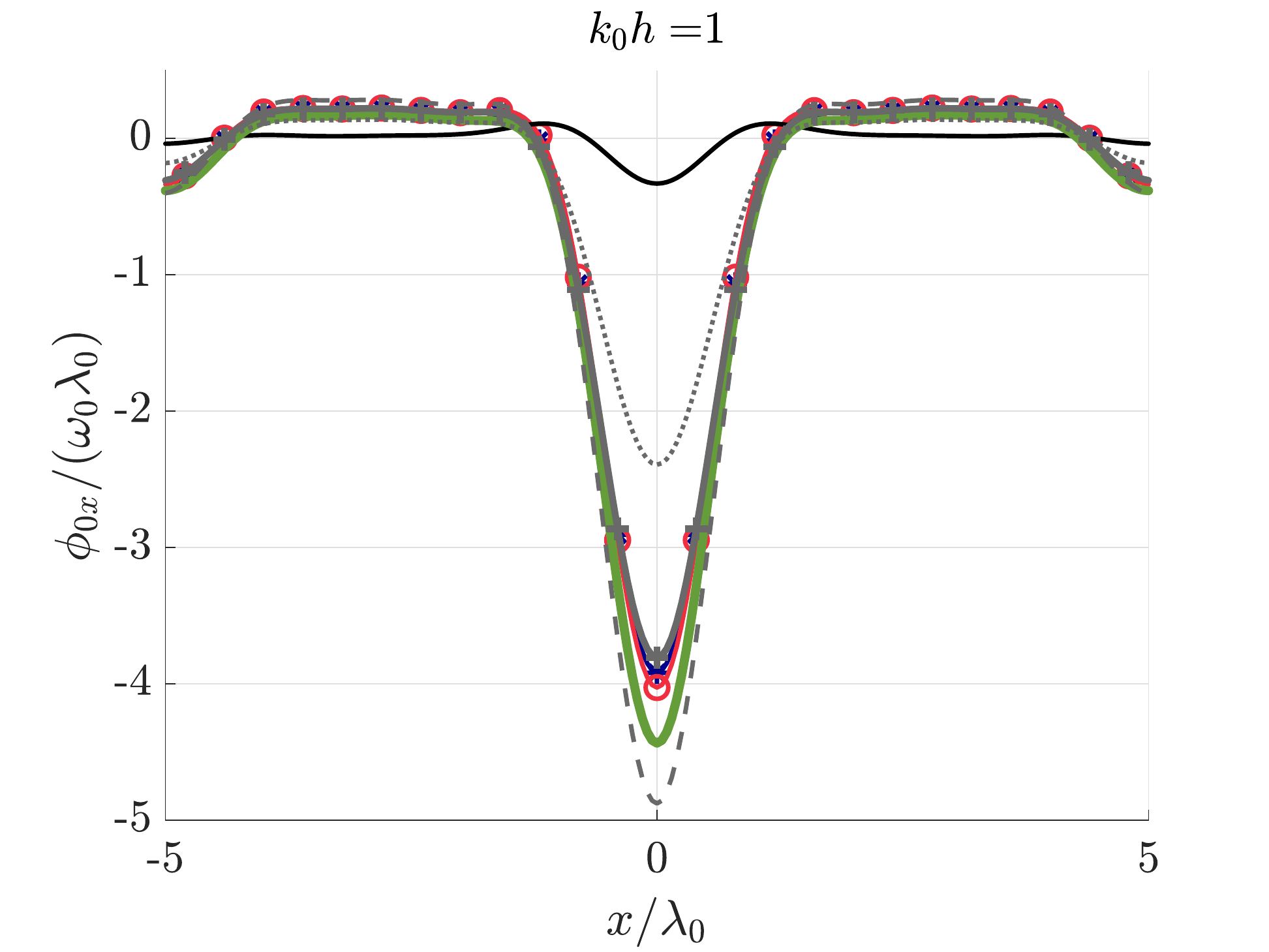}
\includegraphics[width=0.495\textwidth]{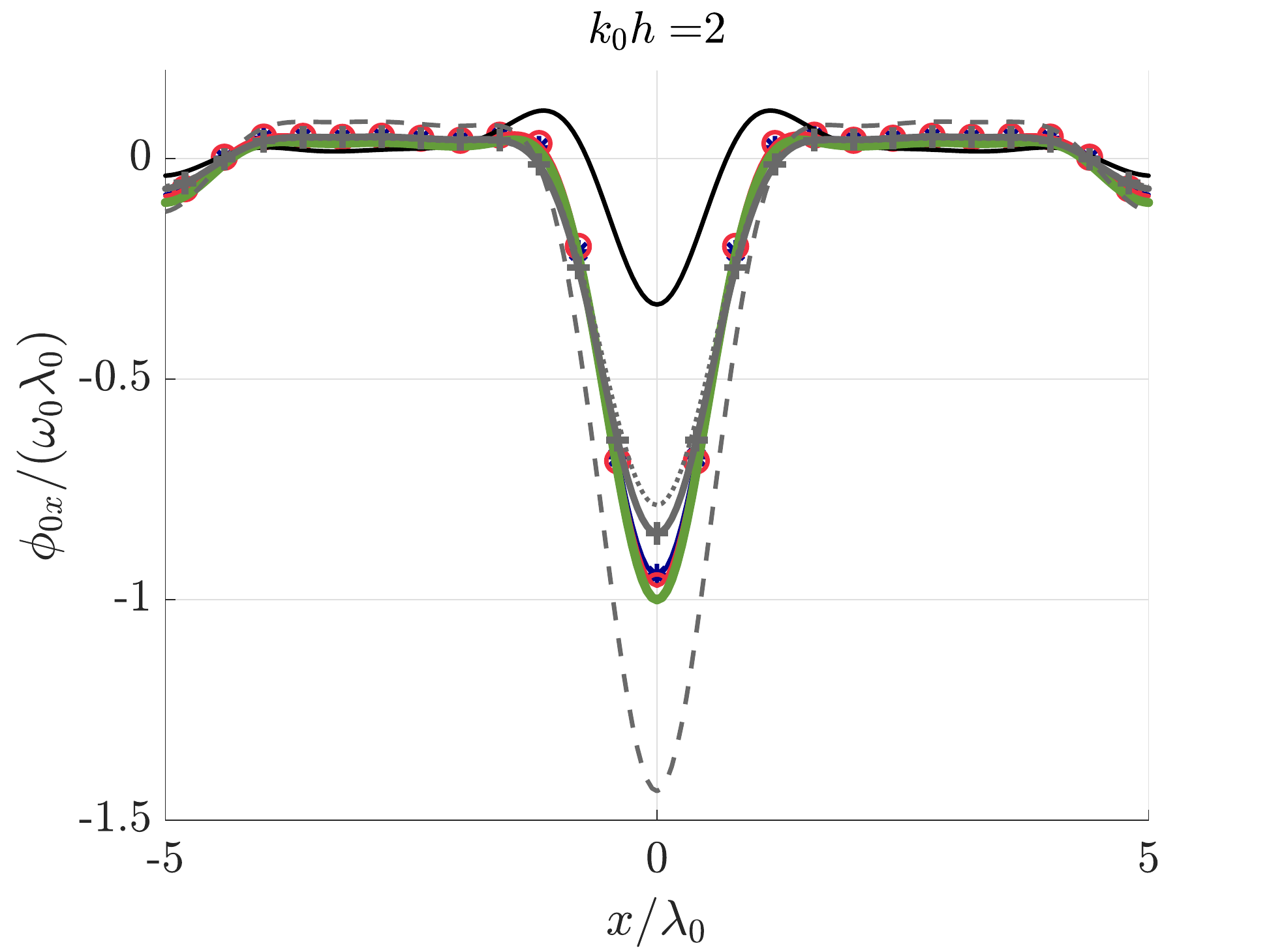}
\includegraphics[width=0.495\textwidth]{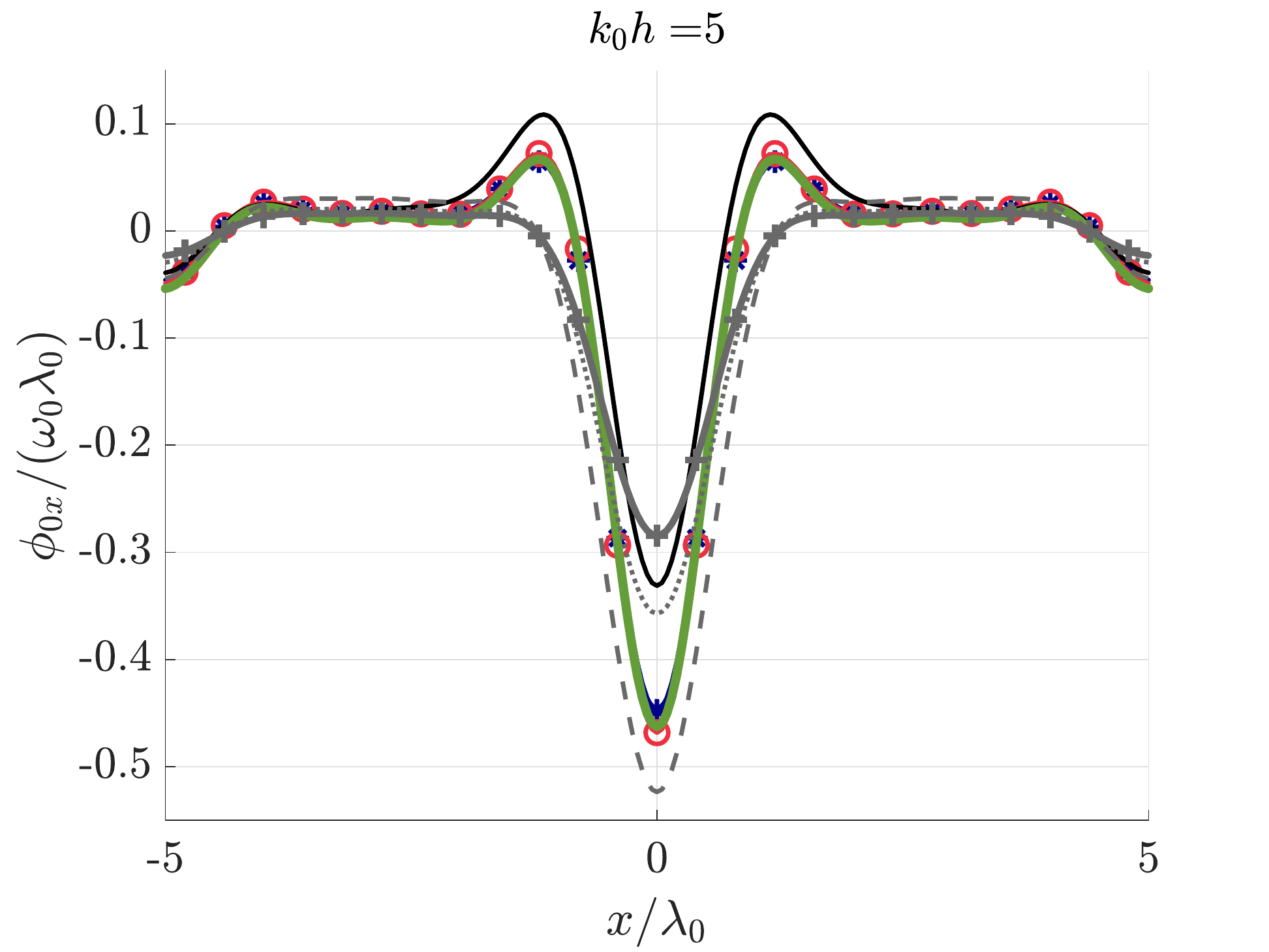}
\includegraphics[width=0.495\textwidth]{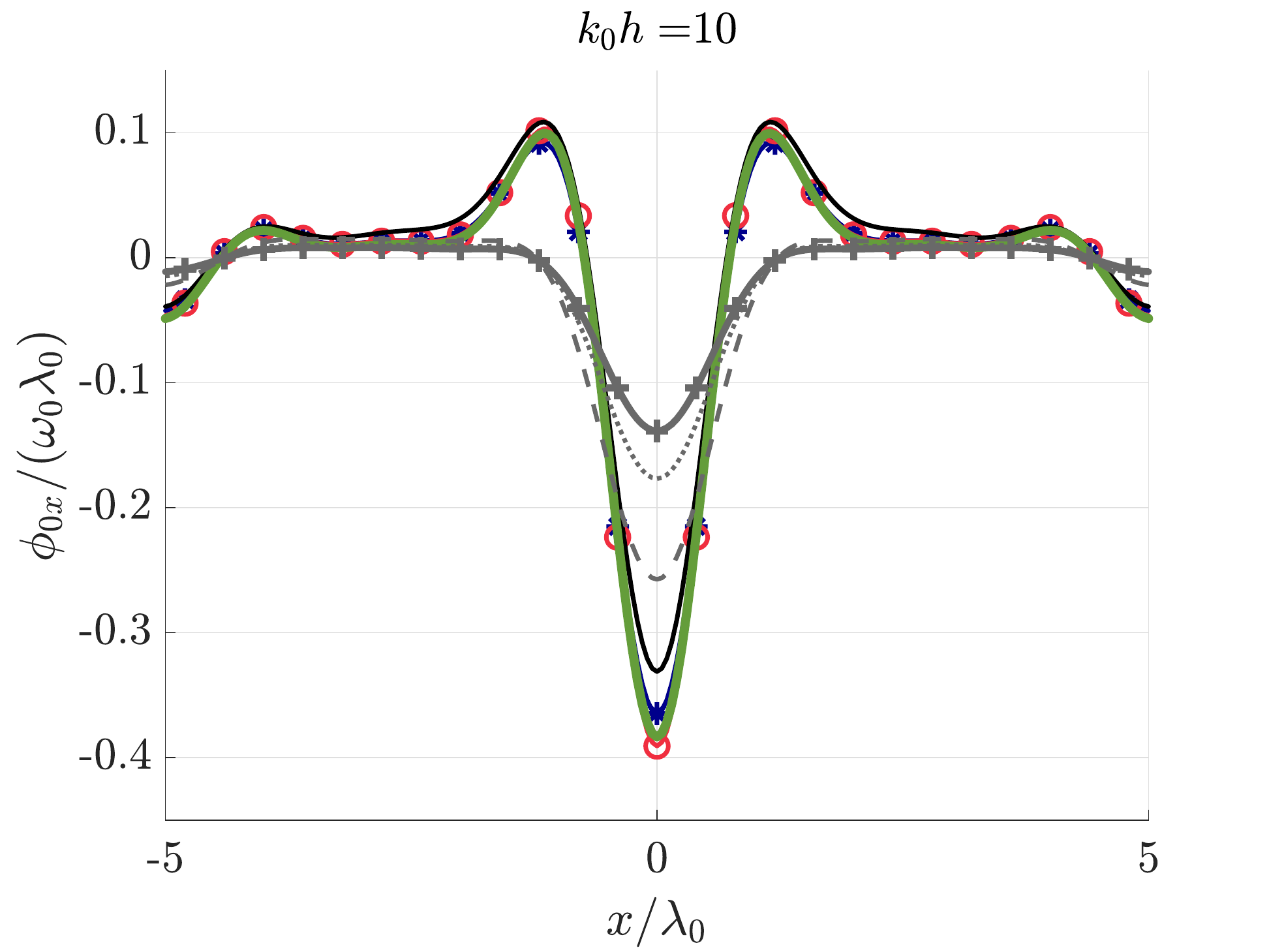}
\includegraphics[width=0.495\textwidth]{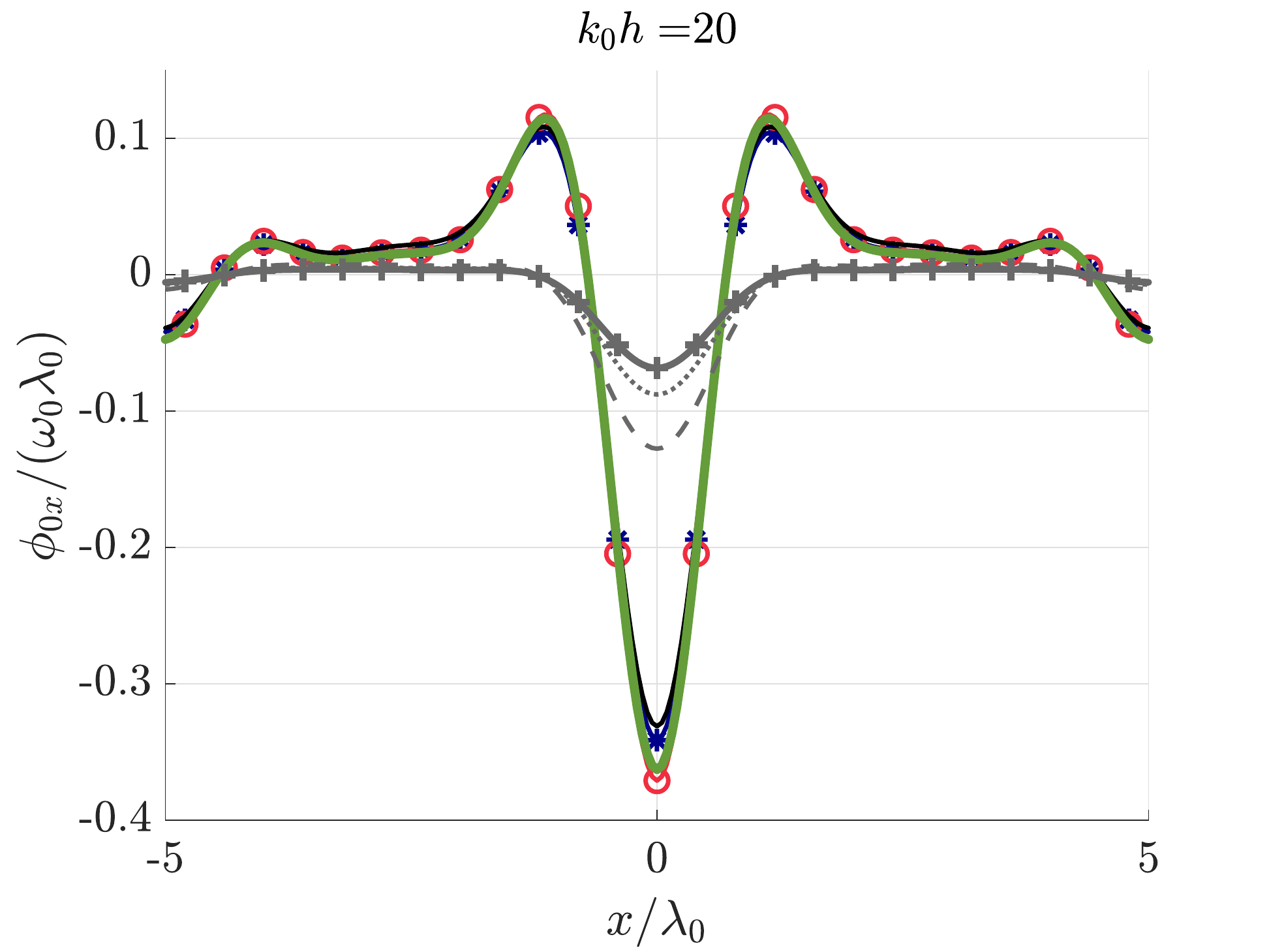}
    \caption{Mean flow $\phi_{0x}$ for $N= 30$ waves focusing at $x=0$ as described by 
    the expressions listed in Table~\ref{tab:table1}, for different values of $k_0 h$.}
\label{fig:focusSL}
\end{figure*}

\begin{figure*}[!]
\includegraphics[width=0.495\textwidth]{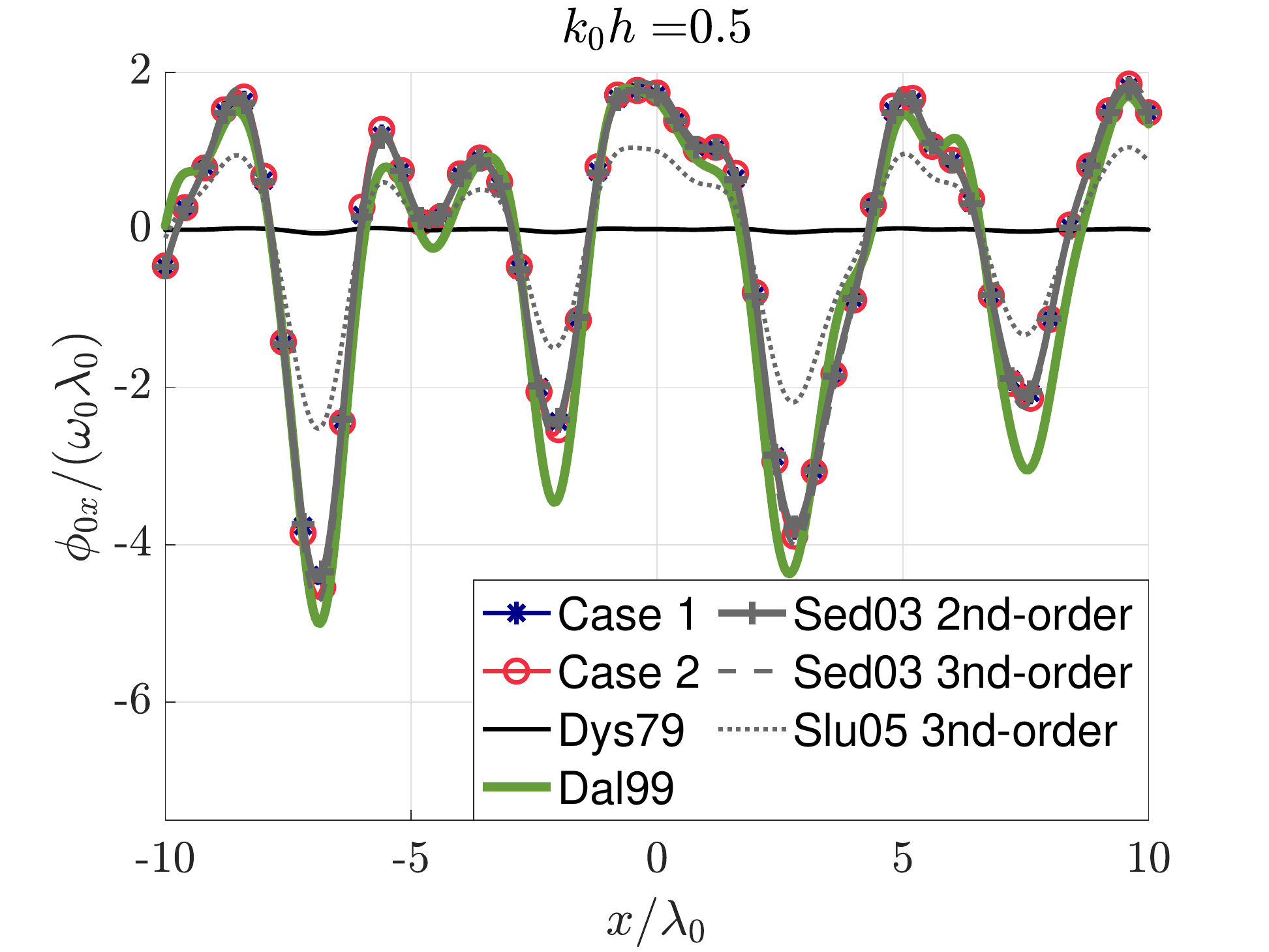}
\includegraphics[width=0.495\textwidth]{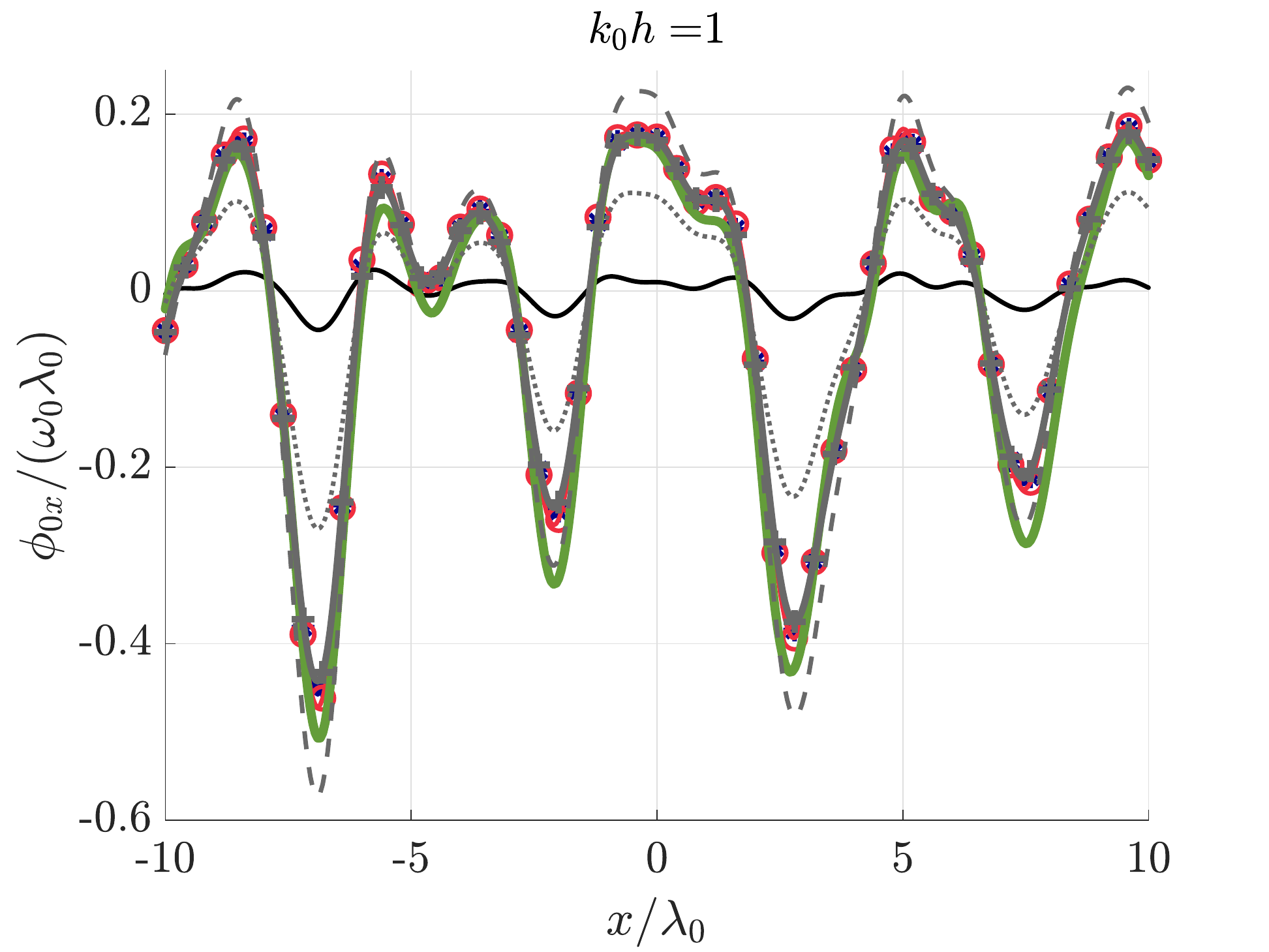}
\includegraphics[width=0.495\textwidth]{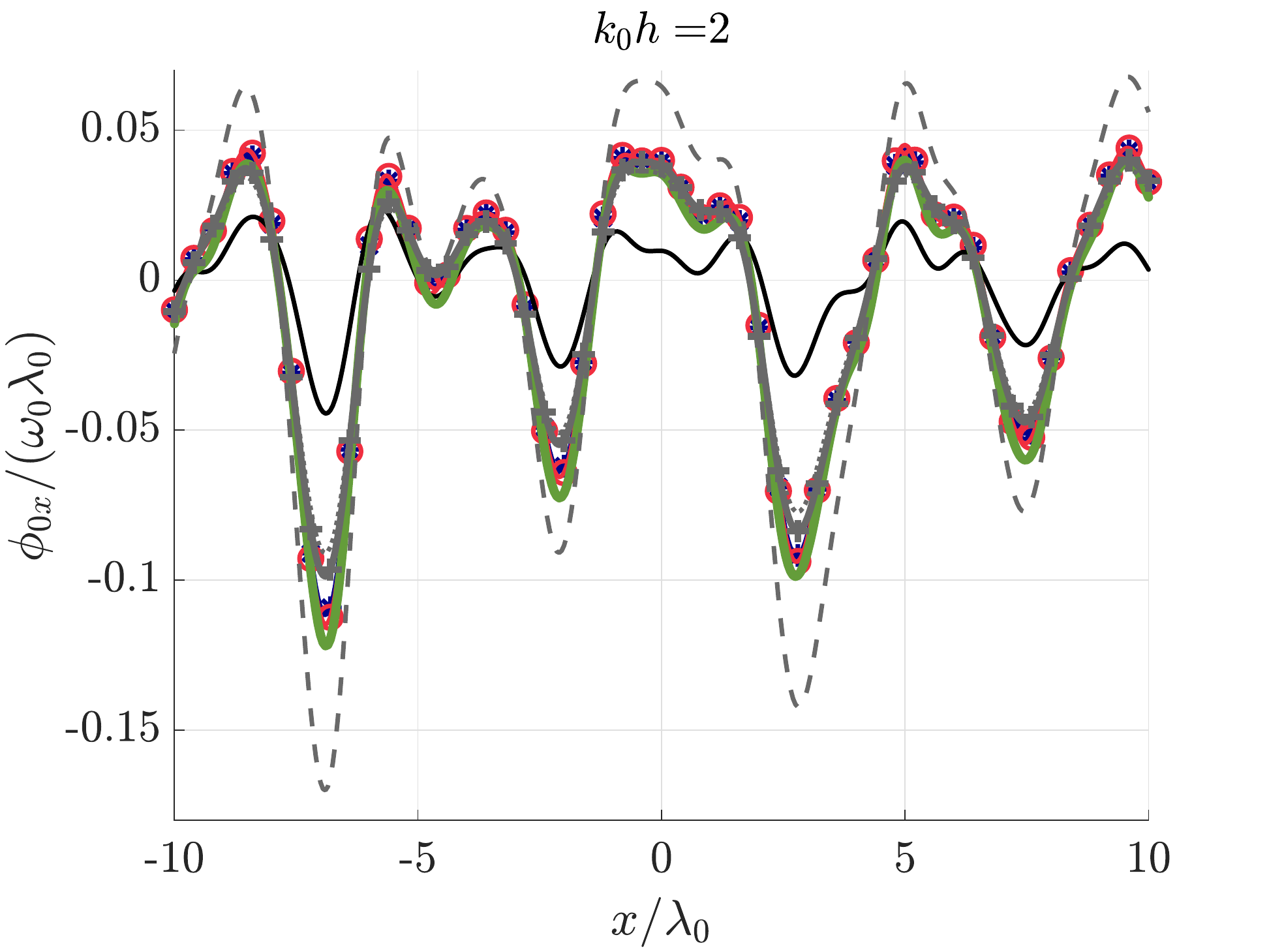}
\includegraphics[width=0.495\textwidth]{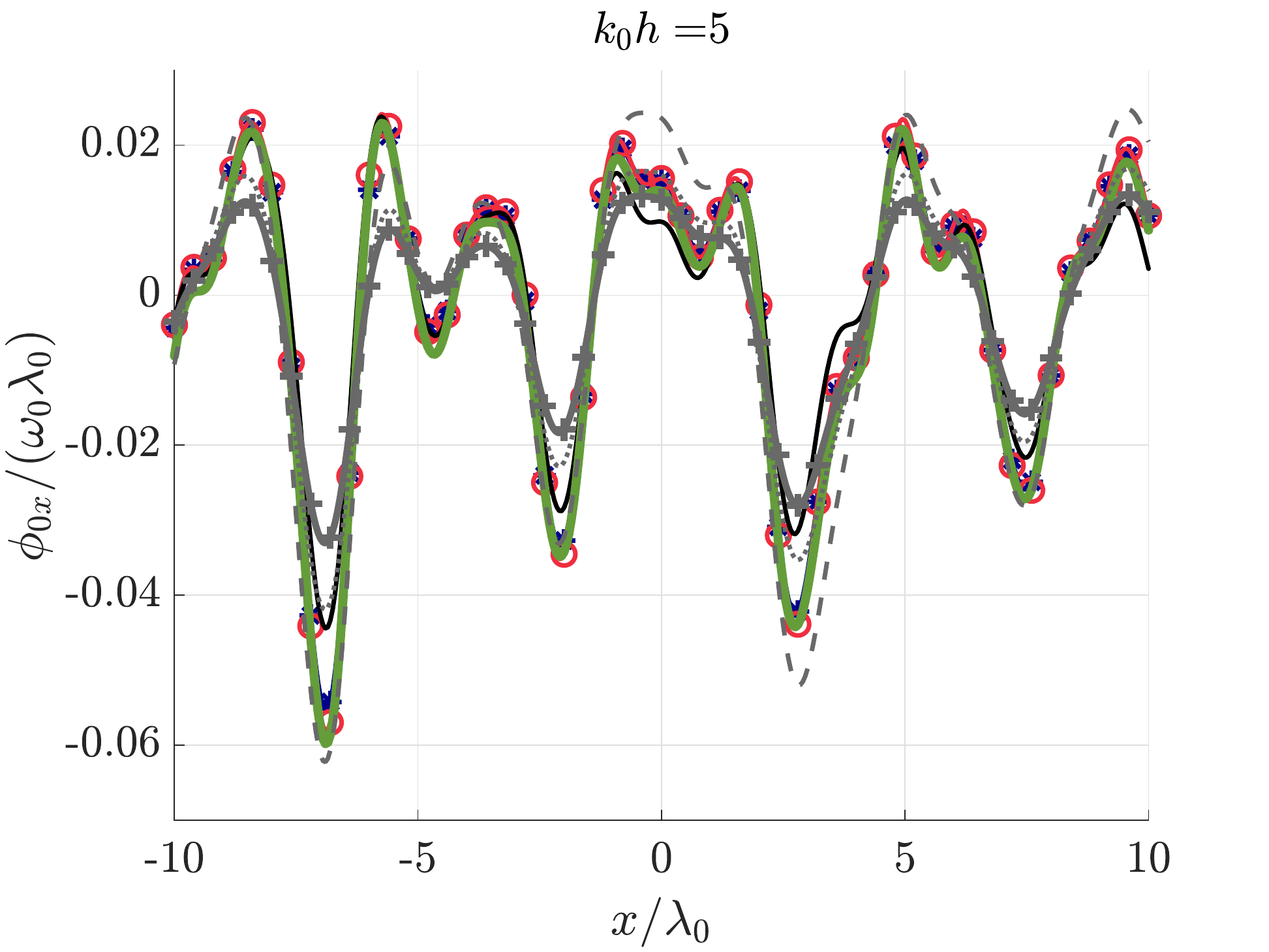}
\includegraphics[width=0.495\textwidth]{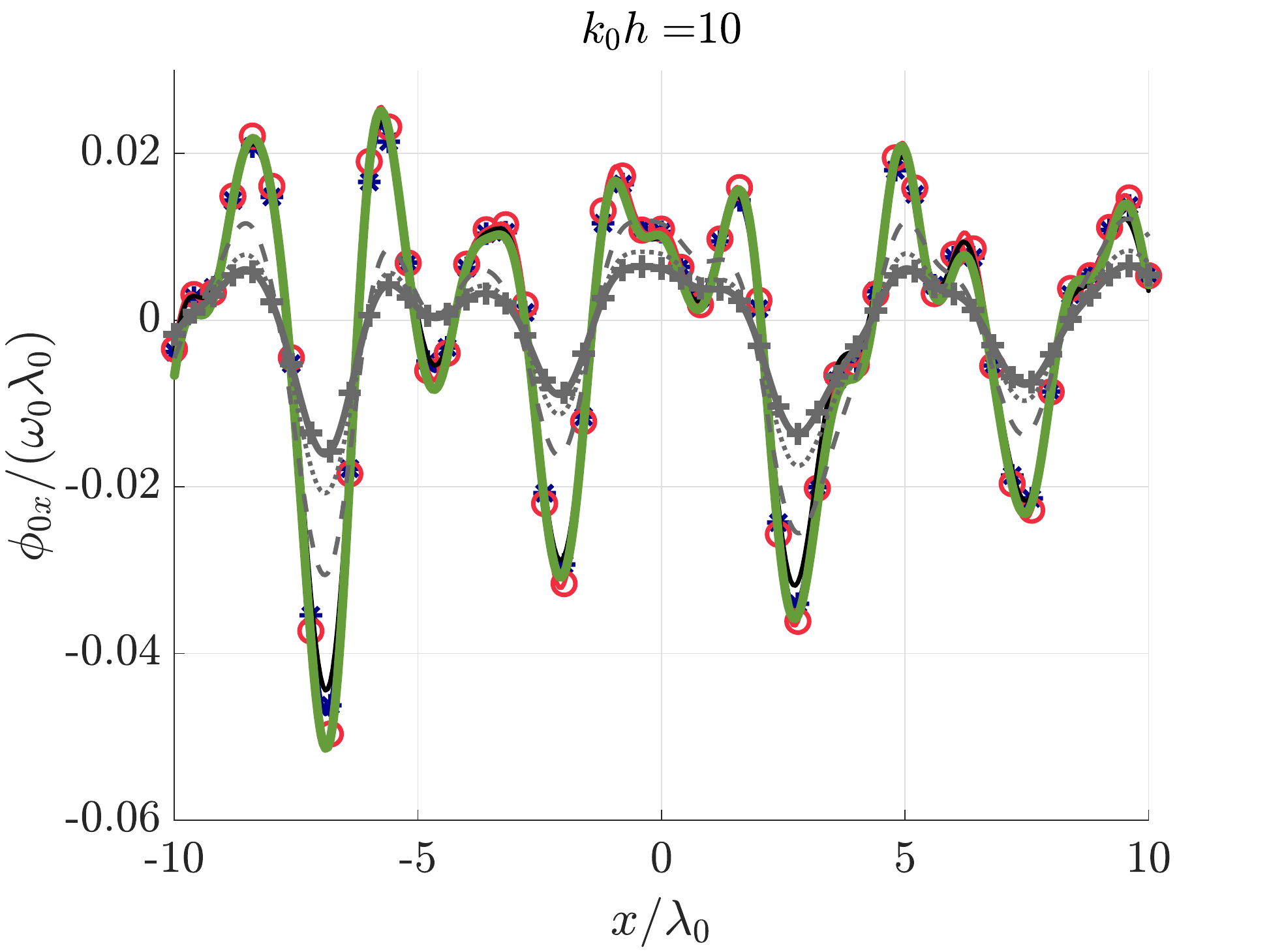}
\includegraphics[width=0.495\textwidth]{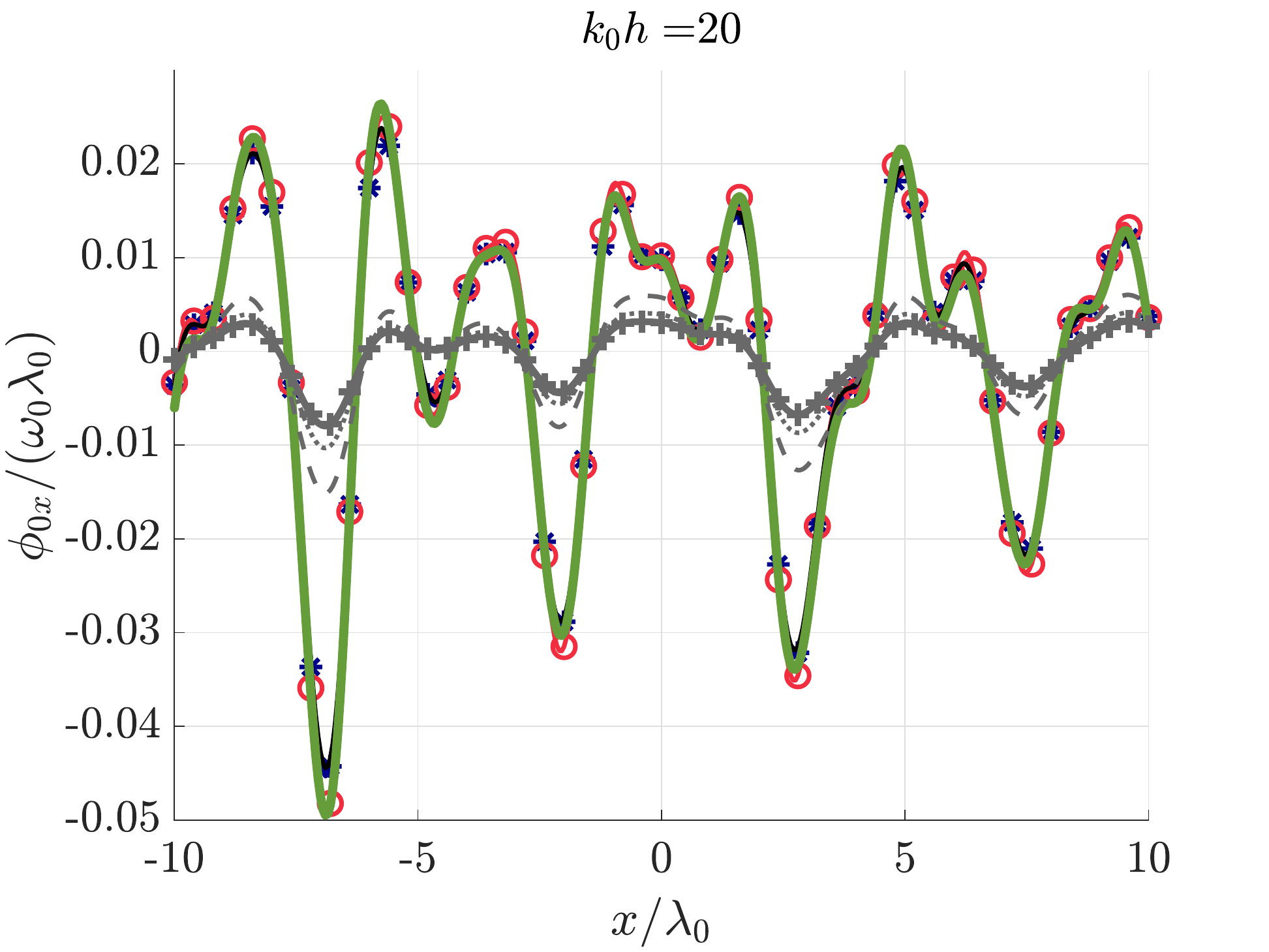}
    \caption{Same as Fig.~\ref{fig:focusSL} for the evolution of $N=30$ waves with random phases.}
\label{fig:nofocusSL}
\end{figure*}

\begin{figure*}[!]
\includegraphics[width=0.49\textwidth]{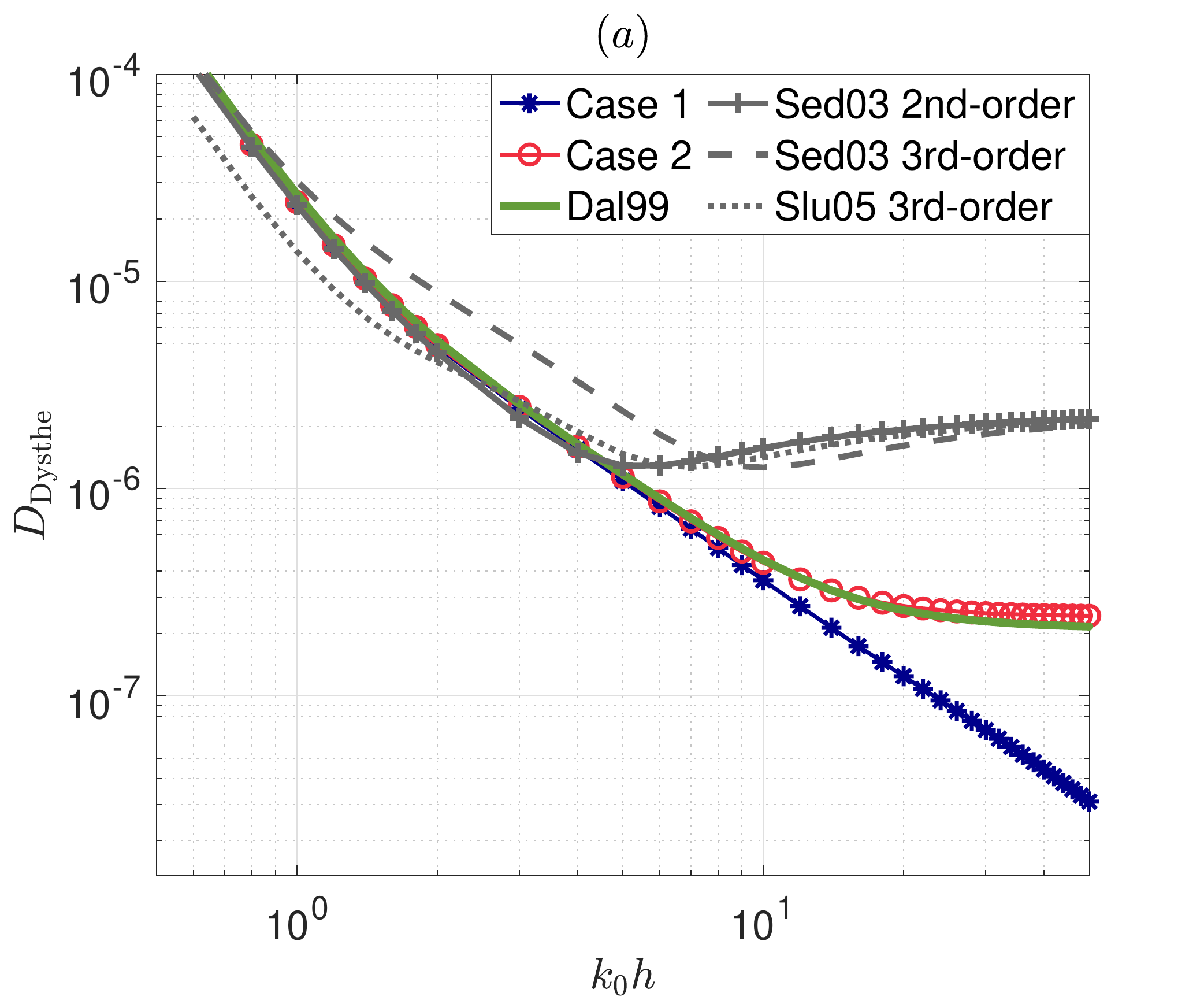}
\includegraphics[width=0.49\textwidth]{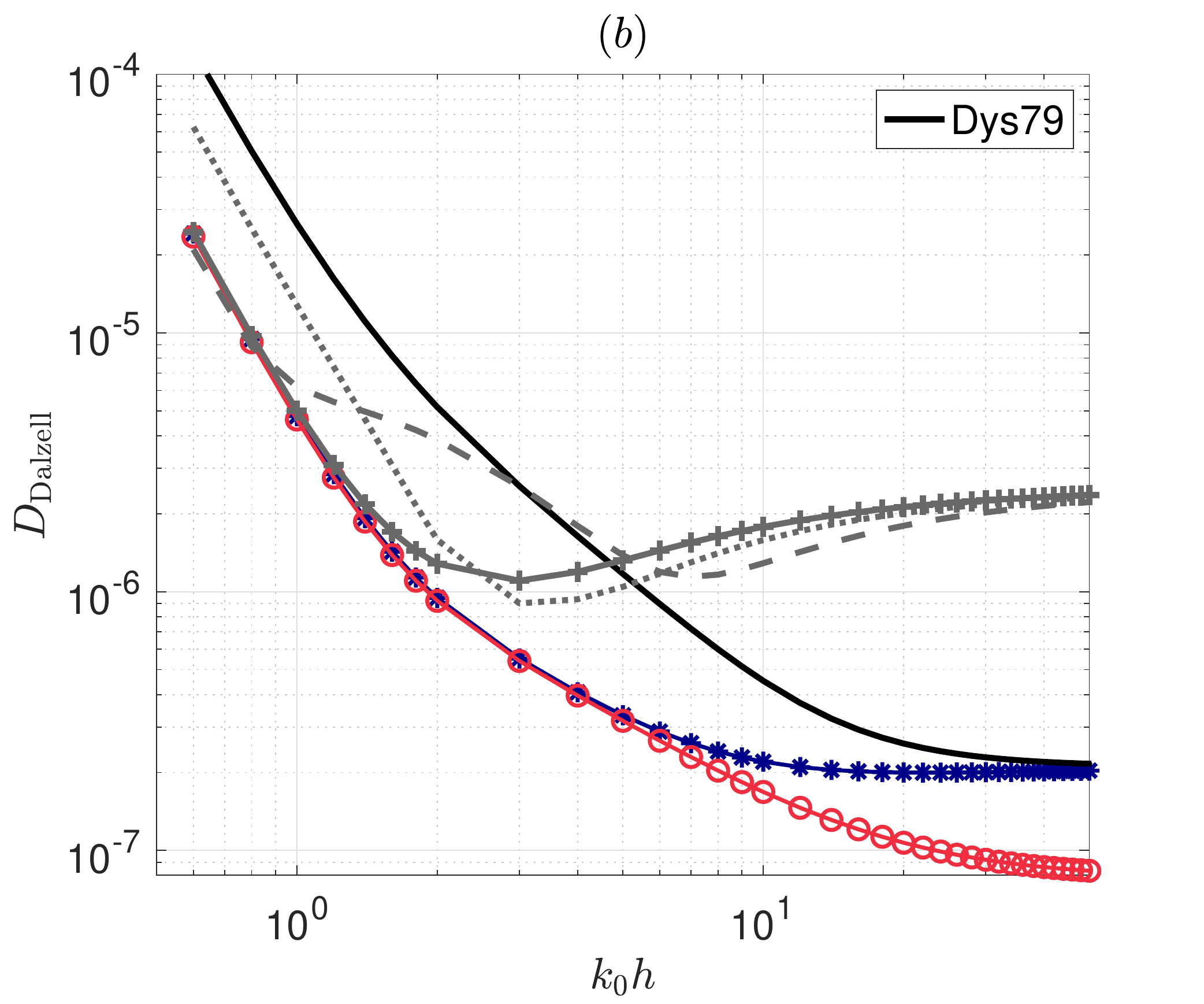}\caption{Deviation operator
with respect to the Dysthe expression ({\it a}) and to the Dalzell one ({\it b}) for the derivative in space $\phi_{0x}$, as defined in the main text, for the different cases listed in Table~\ref{tab:table1}.} 
\label{fig:kh_spacelike}
\end{figure*}

The results are summarised in Fig.~\ref{fig:focusSL} for waves focusing at $x=0$, and in Fig.~\ref{fig:nofocusSL} for the case of waves superposition with random phases. 
In both cases, we see that the horizontal velocity calculated by 
Eq.~(\ref{eq:smallk}) (gray line),  which corresponds to the second-order expression in Sed03, reproduces well the Dalzell waveform (green line) only for $k_0 h \le 2$, and cannot be applied in deep water where it gives a waveform amplitude much smaller than Dysthe's result (black line). 
On the contrary, Dysthe's expression should not be used for $k_0h < 10$ as in our considered cases, especially for large amplitude waveforms, where it does not attain Dalzell's accuracy.
Although an elementary consideration of the dispersion relation suggests that $k_0h\approx5$ should be indistinguishable from $k_0h\to\infty$, the dynamics of the mean flow is consistently different from this limit up to $k_0h=10$.
The expressions for the horizontal velocity given by {\bf Case 1} [Eq.~(\ref{eq:IWspace})] and {\bf Case 2} [Eq.~(\ref{eq:cal2})] behave in a similar way at all depths, providing in general an approximation that corresponds well to the Dalzell solution in intermediate waters, and to Dysthe in deep waters. It is also important to note that the expressions provided in Sed03 and Slu05 
at third-order in steepness [Eqs.~(\ref{eq:mean4order}) and (\ref{eq:mean4orderSLUN}), respectively], give different results and do not correspond to the other models at any depth.   
 
 More detailed features can be deduced from Fig.~\ref{fig:kh_spacelike}, where the deviation operator with respect to  the Dysthe expression, defined as $D_{\rm{Dysthe}} = N^{-1}\int (\phi_{0x}-\phi_{0x}^{\rm{Dysthe}})^2 \,dx$, and to the Dalzell one, $D_{\rm{Dalzell}} = N^{-1}\int (\phi_{0x}-\phi_{0x}^{\rm{Dalzell}})^2\, dx$, with $\phi_{0x} = \partial \phi_0/\partial x$, are shown as a function of the non-dimensional water depth $k_0 h$. The integral is calculated over $L= 80\lambda_0$, and the normalization coefficient is $N = (\omega_{0DW} \lambda_0)^{2}L$, with $\omega_{0DW}$ calculated in deep water (DW). 
 The Stokes series expansion of the velocity potential (and thus Dalzell's model) converges in shallow water~\cite{rahman1970} if $3\Eps/(2k_0h)^3 \ll 1$, thus at low Ursell number, which implies in our case $k_0h \gg 0.48$. This region is excluded in Fig.~\ref{fig:kh_spacelike}. 
It can be seen that the expression for the horizontal velocity given by {\bf Case 2} [Eq.~(\ref{eq:cal2}] is accurate at second-order at all depths, almost superposing to the second-order Dalzell solution, while the one  given by {\bf Case 1} [Eq.~(\ref{eq:IWspace})] is the only expression that consistently converges to the Dysthe in the deep-water limit expression, and is equivalent to {\bf Case 2} for $k_0 h<5$.
Note that the third-order corrections provided in Sed03 [Eq.~(\ref{eq:mean4order})] and in Slu05 [Eq.~(\ref{eq:mean4orderSLUN})] are different, and do not extend the validity to deeper water regimes of the second-order expression [Eq.~(\ref{eq:smallk})], which is accurate for $k_0h<2$ (see 
Fig.~\ref{fig:kh_spacelike}b). This raises a warning on the quantitative accuracy of these third-order expressions. 

\section{Fourth-order equation: propagation in space}
\label{timelike}

\begin{figure}[t!]
 \centering
\includegraphics[width=0.49\textwidth]{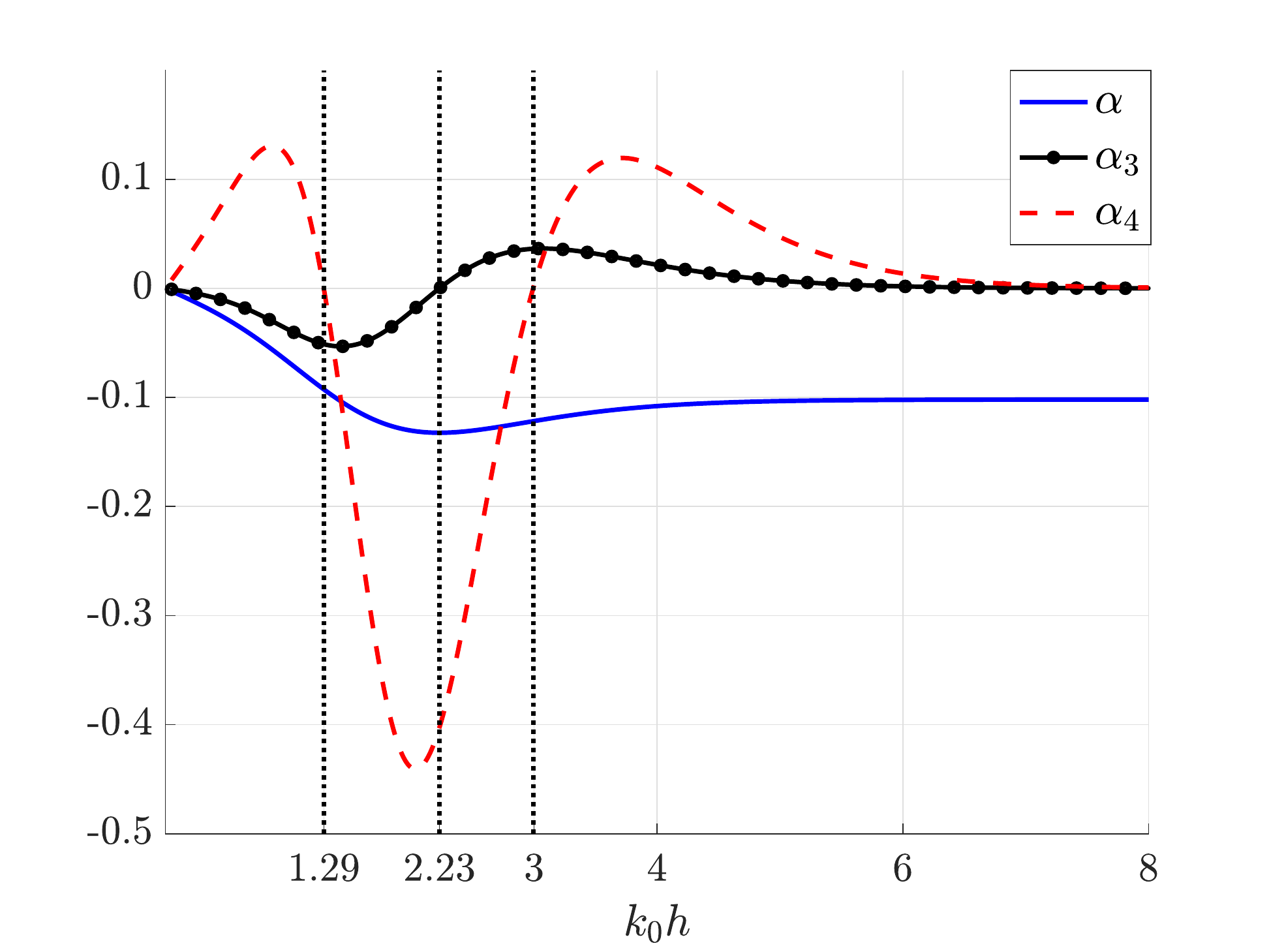}
\caption{Dispersion coefficients (for $\omega_0 = 1$~Hz). Zero crossings are marked on the horizontal axis. 
}
\label{fig:disp}
\end{figure}

\begin{figure*}[th!]
    \centering
\includegraphics[width=0.495\textwidth]{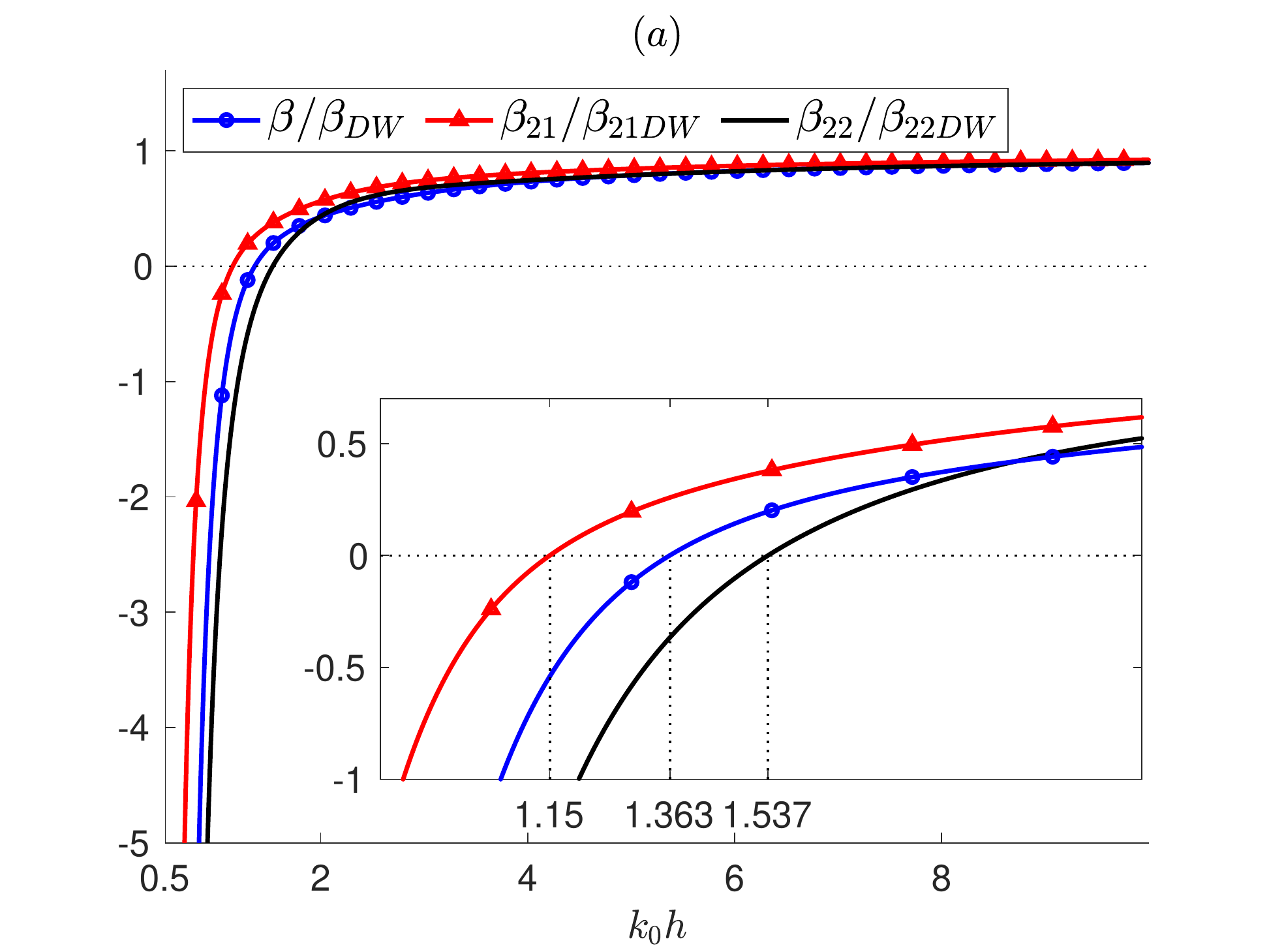}
\includegraphics[width=0.495\textwidth]{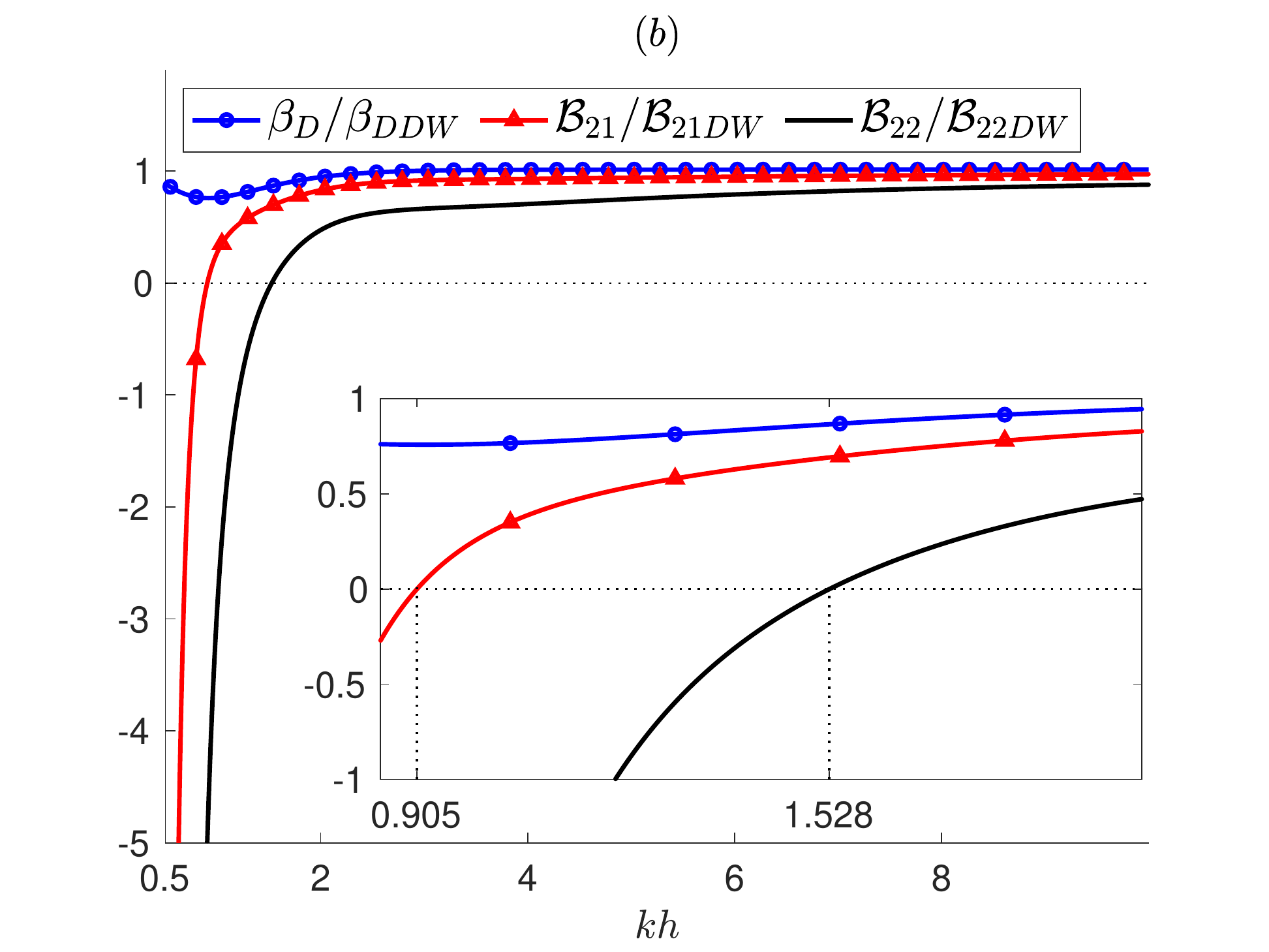}
  \caption{Nonlinear coefficients ({\it a}) in Eq.~(\ref{eq:HONLStime}),  and ({\it b}) in Eq.~(\ref{eq:HONLStimeFINAL}), normalized with respect to their deep-water values (in Dysthe equation, Eq.~(\ref{eq:DystheTimeLike})).  
  Zero crossings are marked in the insets. 
    }
\label{fig:nonlin}
\end{figure*}

In order to transform Eq.~(\ref{eq:HONLSspace}) to an expression describing the propagation in space, that is necessary to describe the nonlinear evolution of waves in a laboratory flume, a change of variables is needed
\begin{equation}
\tau = \Eps\left(t-\frac{x}{c_\mathrm{g}}\right);\; \xi = \Eps^2 x.
\end{equation}
Through this transformation, the third-order terms are only subject to a rescaling,
\begin{equation}
\alpha = \frac{\hat\alpha}{c_\mathrm{g}^3}; \; \beta = \frac{\hat\beta}{c_\mathrm{g}}.
\end{equation}
The higher-order terms are instead dramatically modified. 
Indeed, a mixed derivative appears from the second-order dispersion term, $\partial^2{U}/\partial{x}^{2}$, that reads as
$- (\omega''/c_\mathrm{g})\partial^2 U/\partial x\partial t$. 
In the multiscale spirit, the time-like NLS ({\it i.e.}, terms at third-order in steepness)
is used to estimate this term, which results in corrections at fourth-order.
The resulting time-like form of the evolution equation does not appear explicitly in the literature and is given by (setting $\Eps =1$, and changing the notation as $\tau \to t$ and $\xi \to x$):  
\begin{widetext}
\begin{equation}
			i\pder{U}{x} + \alpha\pdern{U}{t}{2} \underbrace{- \beta|U|^2U}_\text{incl.~Mean Flow} =
			 -i\alpha_3 \pdern{U}{t}{3}
			 \underbrace{+ i\beta_{21} |U|^2 \pder{U}{t}
			  +  i\beta_{22} U^2 \pder{U^*}{t}}_\text{incl.~Mean Flow},
			   \label{eq:HONLStime}
\end{equation}
whose dispersion and nonlinear coefficients are given in App.~\ref{app:A} and \ref{app:B}, and their dependence on $k_0 h$ is shown in Figs.~\ref{fig:disp} and \ref{fig:nonlin}.
High-order dispersion terms~\cite{[{In particular, the fourth-order dispersion term 
$\alpha_4 \partial^4 U/\partial t^4$ (to be added on the r.h.s.~of Eq.~(\ref{eq:HONLStime})) is generally used to cancel out resonances that can make the model numerically unstable, as discussed in~}]hara_mei_1991} are easily included up to arbitrary order 
following Refs.~\onlinecite{TrulsenDispersion2000,EeltinkKurtosis2019}, reducing the constraints in bandwidth of the original Dysthe equation.

Using the relation~\cite{Sedletsky2003} 
\begin{equation}
\pder{\phi_{01}}{t} = -c_g  \pder{\phi_{01}}{x} 
\label{eq:cgphi0}
\end{equation} 
Eq.~(\ref{eq:HONLStime}) can be rewritten in the equivalent form:
\begin{equation}
			i\pder{U}{x} + \alpha\pdern{U}{t}{2} - \beta_D |U|^2U =
			 -i\alpha_3 \pdern{U}{t}{3}
			  + i\mathcal{B}_{21} |U|^2 \pder{U}{t}
			   +  i\mathcal{B}_{22} U^2 \pder{U^*}{t} 
			\underbrace{- \frac{\mu_g k_0}{4\sigma c_g^2 } U\pder{\phi_0}{t}}_\text{Mean Flow}
\label{eq:HONLStimeFINAL}
\end{equation}
\end{widetext}
where $\beta_D =\hat \beta_D/c_g$, 
$\mathcal{B}_{21} = \omega_0 k_0 \tilde Q_{41}/c_g^2 - 4\alpha\beta_D c_g$,
$\mathcal{B}_{22} =  \omega_0 k_0 \tilde Q_{42}/c_\mathrm{g}^2-2\alpha\beta_D c_\mathrm{g}$, with 
$\hat\beta_D$, $\alpha$, $\alpha_3$ given in App.~\ref{app:A}, and $\tilde Q_{41}$, $\tilde Q_{42}$ in App.~\ref{app:B}. The dependence of the nonlinear coefficients on $k_0 h$ is illustrated in Fig.~\ref{fig:nonlin}. 
Note that $\beta, \beta_{21}, \beta_{22}, \mathcal{B}_{21}$ and  $\mathcal{B}_{22}$ diverge quite strongly for $k_0h\to 0$, {\it i.e.} particularly in the defocusing regime. 
That said and as can be also analytically verified, all coefficients correctly reproduce their deep-water limit at $k_0h \to \infty$. 
  
\subsection{Mean flow in time-like equations}

At main order in steepness, Eqs.~(\ref{eq:mean4order})  and (\ref{eq:cgphi0}) imply \begin{equation}
\pder{|U|^2}{t} = -c_g  \pder{|U|^2}{x}
\label{eq:timederiv}
\end{equation}
Thus, $\phi_{01}$ and $|U|^2$ do not depend on $x$ and $t$ separately, but only through their combination $(x-c_g t)$, so that at leading order the Fourier transform in space, ${\cal F}_x$, can be replaced by the Fourier transform in time, ${\cal F}_t$. From 
Eq.~(\ref{HilbertTerm}), in the deep-water limit it is possible to simply exchange the derivative with respect to space with that with respect to time and vice versa, since the Hilbert transform of the derivative is the derivative of the Hilbert transform, {\it i.e.} these two linear operators commute. As such:
\begin{equation}
    \pder{\phi_0}{t}=\frac{\omega_0}{2}\mathcal{H}_t
    \left[\pder{|U|^2}{t}\right]
\label{eq:hilbert_timelike}    
\end{equation}
The time-like Dysthe equation is given by the deep-water limit of Eq.~(\ref{eq:HONLStimeFINAL}) and reads, using the above expression for the mean flow term~\cite{TrulsenDysthe1997,kitShemer2002,OnoratoOsborneEtAl2005,GoulletChoi2011,Eeltink2017}
\begin{widetext}
\begin{equation}
			i\pder{U}{x} - \frac{k_0}{\omega_0^2}\pdern{U}{t}{2} - k_0^3 |U|^2U =
			  2 i \frac{k_0^3}{\omega_0}
     \biggl(4 |U|^2 \pder{U}{t}
			   +   U^2 \pder{U^*}{t}+\underbrace{iU \mathcal{H}_t\left[\pder{|U|^2}{t}\right]}_\text{Mean Flow} \biggr) 
\label{eq:DystheTimeLike}
\end{equation}
where $\alpha_3 \to 0$, since $k \propto \omega^2$, and $\hat\beta_{21} \to \frac{3}{2}\omega_0 k_0$, $\hat\beta_{22} \to \frac{1}{4}\omega_0 k_0$, $\beta_{21} \to  8 k_0^3/\omega_0$, and $\beta_{22} \to  2k_0^3/\omega_0$. 

For the last term in Eq.~(\ref{eq:HONLStimeFINAL}), the Sed03 expression, as in Eq.~(\ref{eq:mean4order}), can be written in terms of the time derivative, using Eq.~(\ref{eq:cgphi0}):  
\begin{eqnarray}
&&\pder{\phi_0}{t} = -c_g  \frac{\omega_0}{2} \frac{k_0 \mu_g}{\sigma \nu}  |U|^2 \nonumber \\ 
&-& i \frac{4\omega_0 \sigma}{\nu} \tilde q_{40S} \left(U \pder{U^*}{t}- U^* \pder{U}{t}\right).  
\label{eq:mean4orderTime}
\end{eqnarray}
However, this expression goes to zero in the deep-water limit, and thus it does not converge to the Dysthe mean flow given in Eq.~(\ref{eq:hilbert_timelike}).

Replacing Eq.~(\ref{eq:cgphi0}) in the Laplace equation and the surface boundary conditions, and repeating the same steps of  Sec.~\ref{subsec:arbitrary}, we finally get the following expressions for the derivative in time of $\phi_0$ for {\bf Case 1} and {\bf 2}, respectively, to be inserted in the evolution Eq.~(\ref{eq:HONLStimeFINAL}): 
\begin{eqnarray}
  \pder{\phi_0}{t} &=& D\mathcal{F}^{-1}_t\Bigg\{\frac{i}{\tanh(\omega h/c_g)} \mathcal{F}_t \left[\pder{|U|^2}{t}\right]\Bigg\},  \,\rm{for~Case~1} 
\label{eq:IWtime_case1}  \\
 \pder{\phi_0}{t} &=& D'\mathcal{F}^{-1}_t\Bigg\{\frac{i}{\tanh(\omega h/c_g)\left[1-c_g \omega/(g\tanh(\omega h/c_g))\right]} \mathcal{F}_t \left[\pder{|U|^2}{t}\right]\Bigg\},   \, \rm{for~Case~2}
\label{eq:IWtime_case2}    
\end{eqnarray}
\end{widetext}

\begin{figure*}[!]
    \centering
\includegraphics[width=0.495\textwidth]{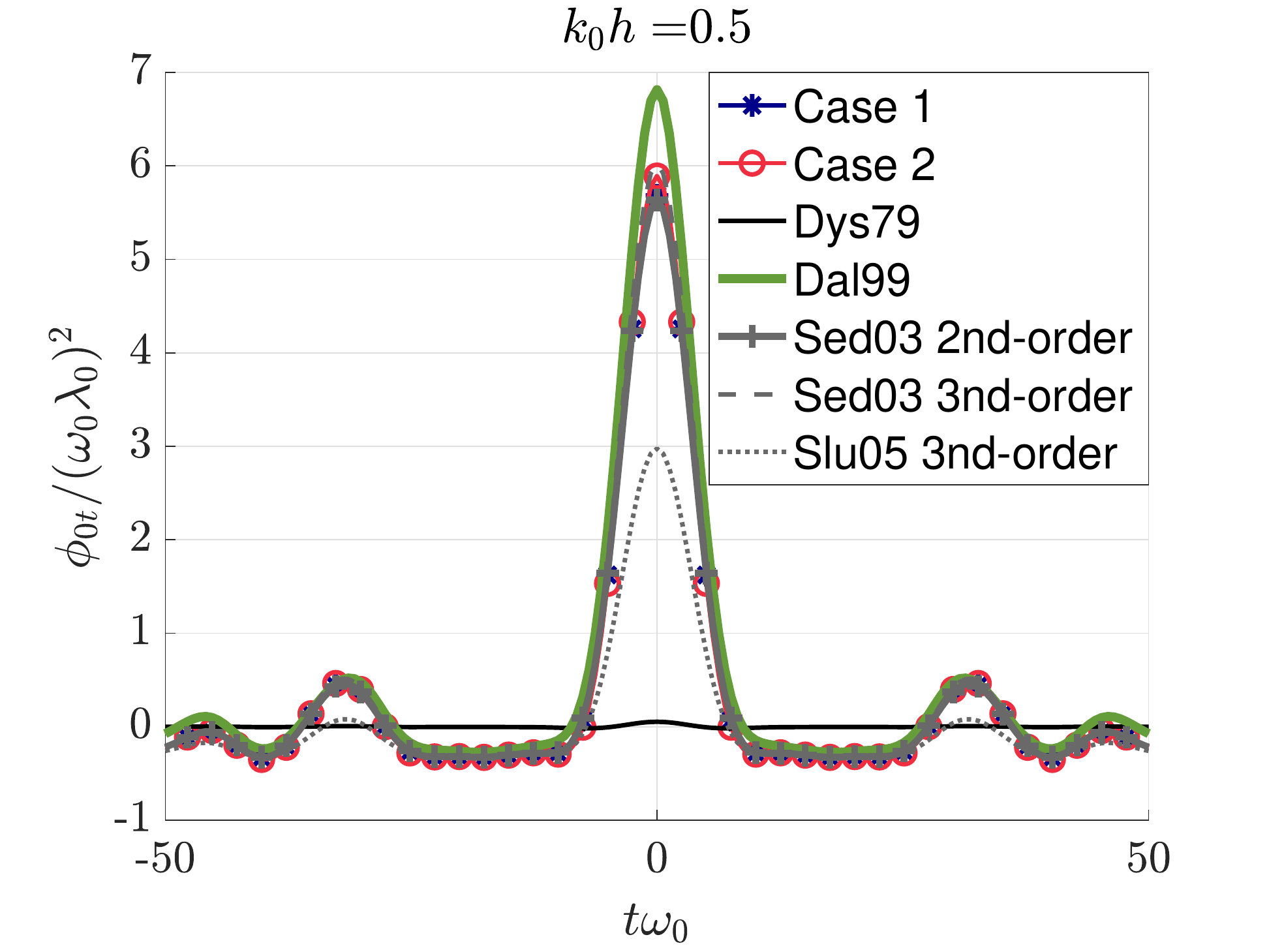}
\includegraphics[width=0.495\textwidth]{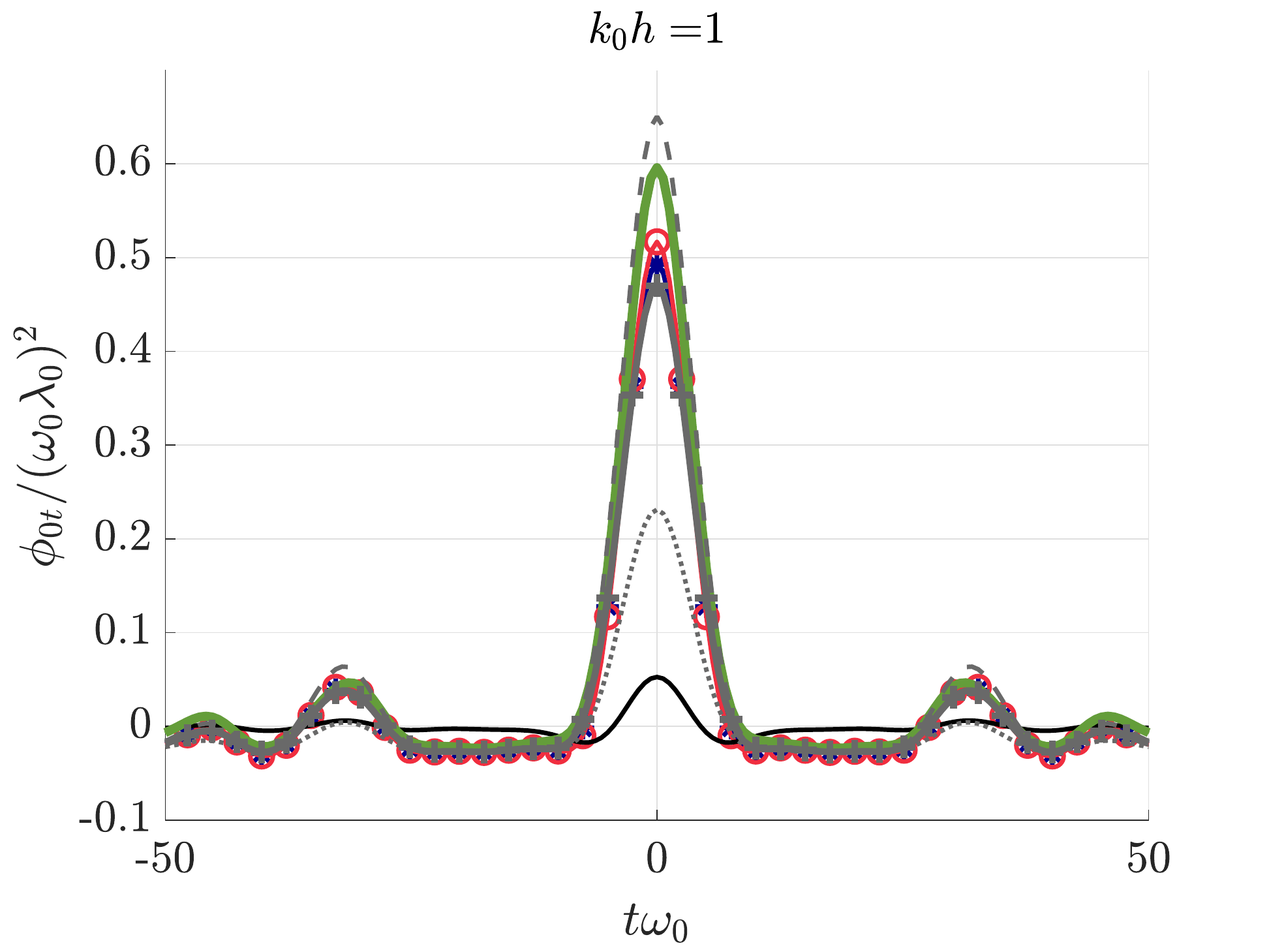}
\includegraphics[width=0.495\textwidth]{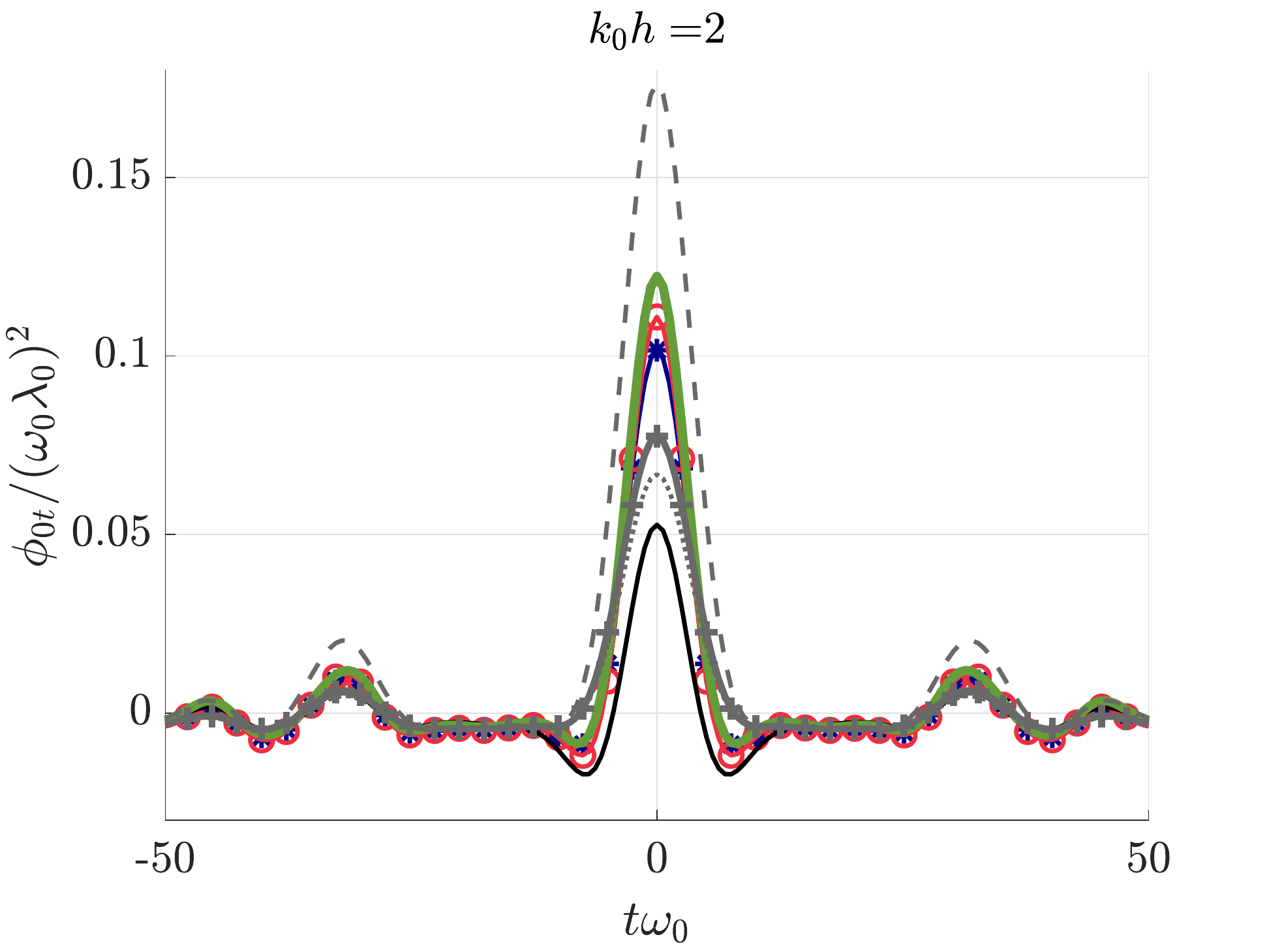}
\includegraphics[width=0.495\textwidth]{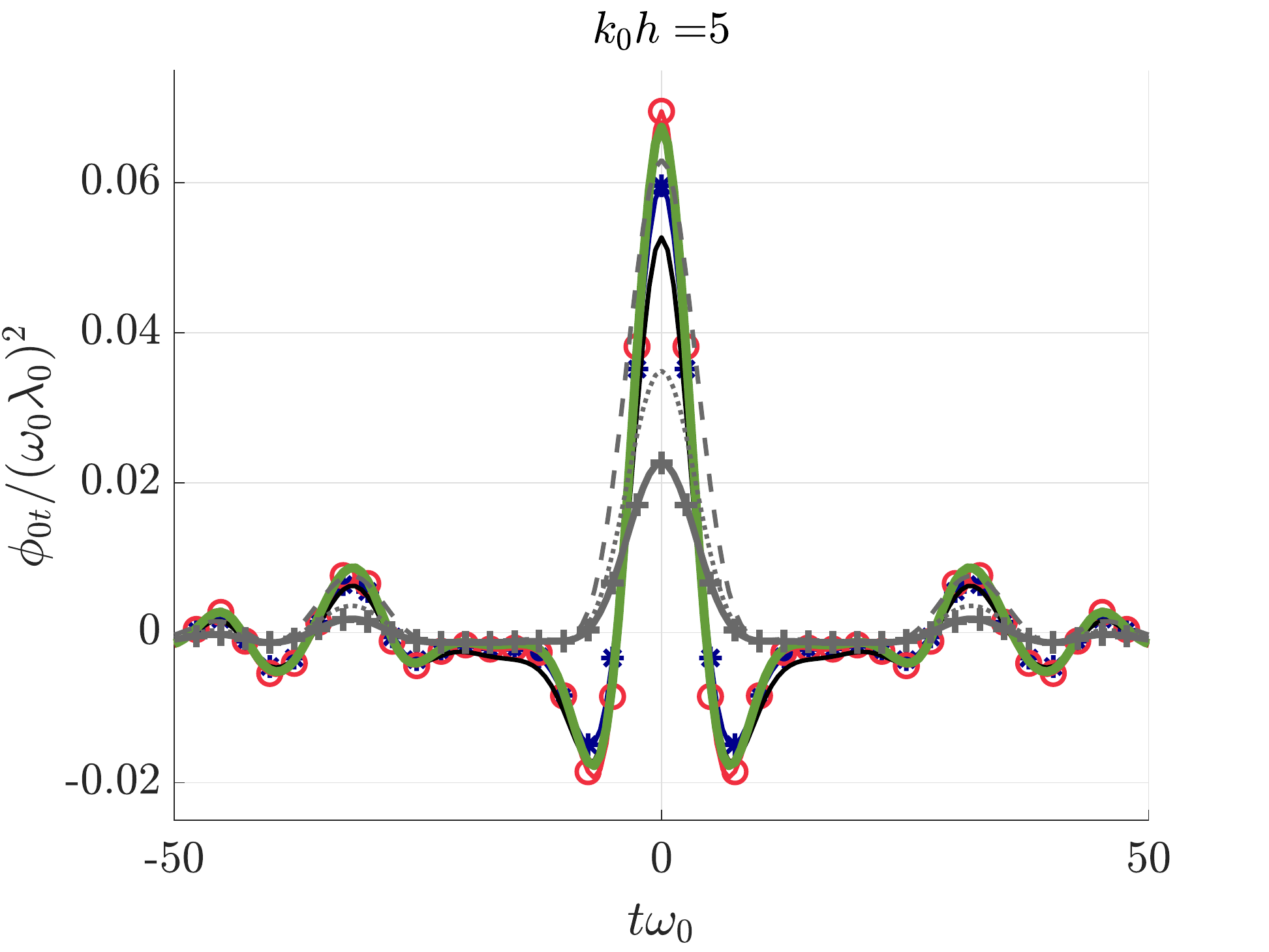}
\includegraphics[width=0.495\textwidth]{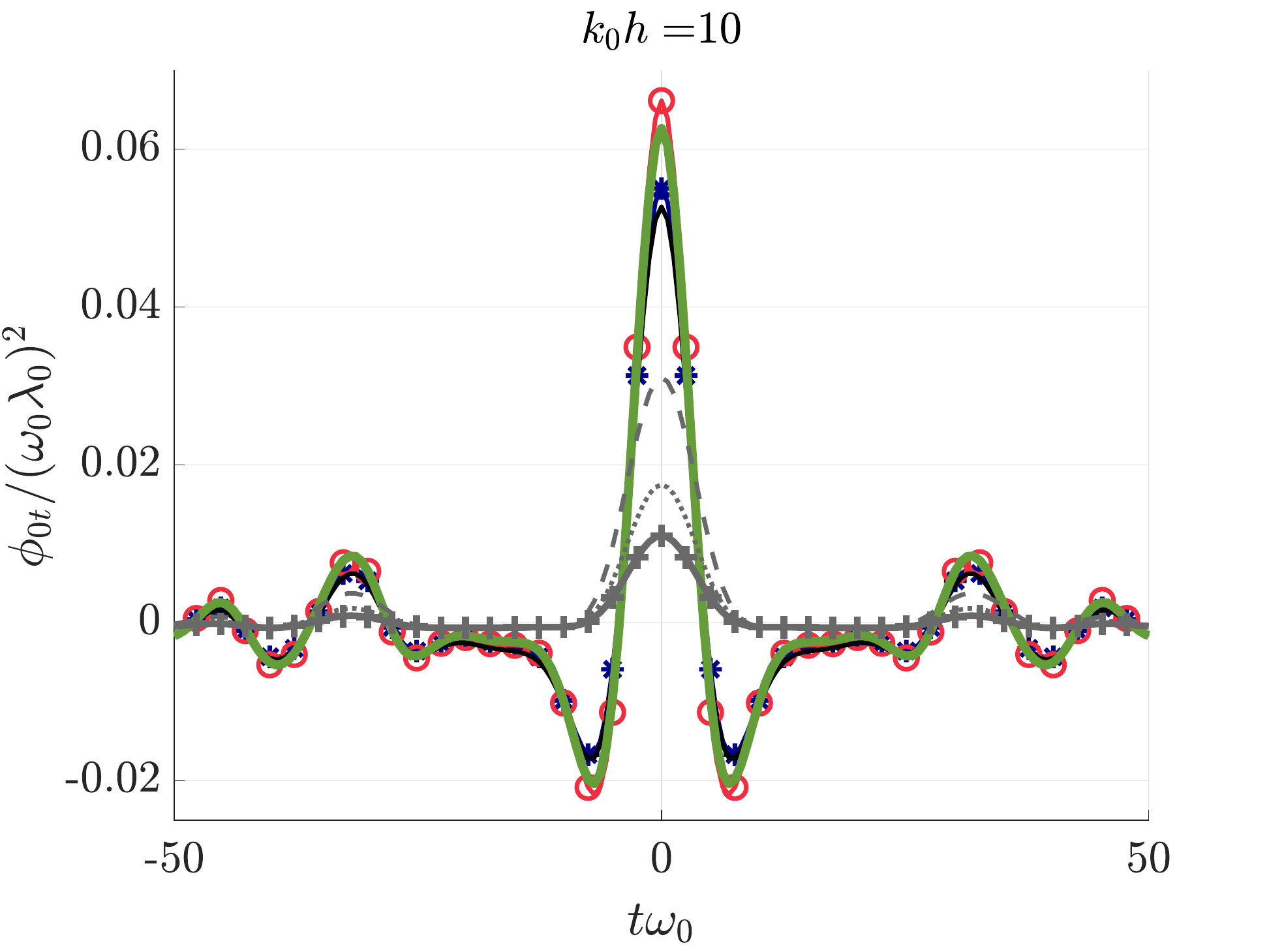}
\includegraphics[width=0.495\textwidth]{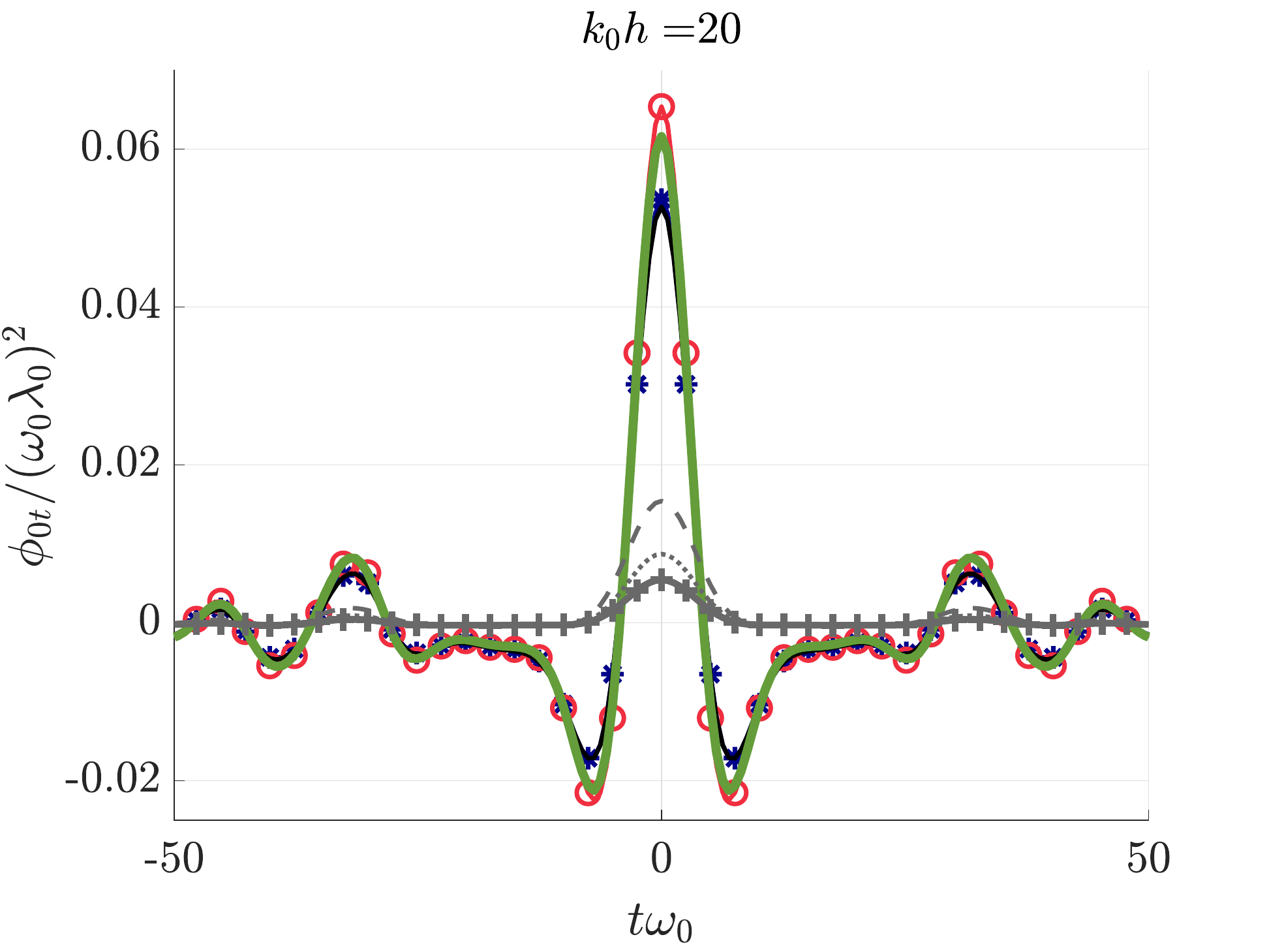}
    \caption{Mean flow $\phi_{0t}$ for N = 30 waves focusing at $t=0$ described by 
    the expressions listed in Table~\ref{tab:table2}, for different values of $k_0 h$.}
 \label{fig:focusTL}
\end{figure*}

\subsection{Numerical comparisons}

\begin{table}
\caption{\label{tab:table2} 
List of mean flow terms in time-like formulation: $\partial \phi_0/\partial t$. }
\begin{ruledtabular}
\begin{tabular}{lll}
{\bf Case 1} & Eq.~(\ref{eq:IWtime_case1}) & 3rd-order, \\ 
&& non-instantaneous \\
{\bf Case 2} & Eq.~(\ref{eq:IWtime_case2}) & 3rd-order, non-inst. \\
Dys79~\cite{Dysthe1979}  & Eq.~(\ref{eq:hilbert_timelike})  & 3rd-order, non-inst. \\
&& deep water \\
Sed03~\cite{Sedletsky2003}  & Eq.~(\ref{eq:mean4orderTime})  & 3rd-order \\
Sed03  & 1st term on  & 2nd-order \\
& r.h.s of Eq.~(\ref{eq:mean4orderTime}) & \\
Slu05~\cite{Slunyaev2005}  & Eq.~(\ref{eq:mean4orderSLUNtime})  & 3rd-order \\
\end{tabular}
\end{ruledtabular}
\end{table}

We now compare the expressions for $\partial \phi_0/\partial t$ listed in Table~\ref{tab:table2} with the sub-harmonic velocity potential $\phi_{20}$ at second-order in steepness and its time derivative calculated using the Dalzell analytical method 
in Refs.~\onlinecite{dalzell1999,LiLi2021}. 

We use the same parameters as in Sec.~\ref{sec:comparisons}, starting in this case by  a Gaussian (amplitude) spectrum $S(\omega)$ peaked at $\omega_0 = 1$~Hz. 

\begin{figure*}[ht!]
    \centering
\includegraphics[width=0.49\linewidth]{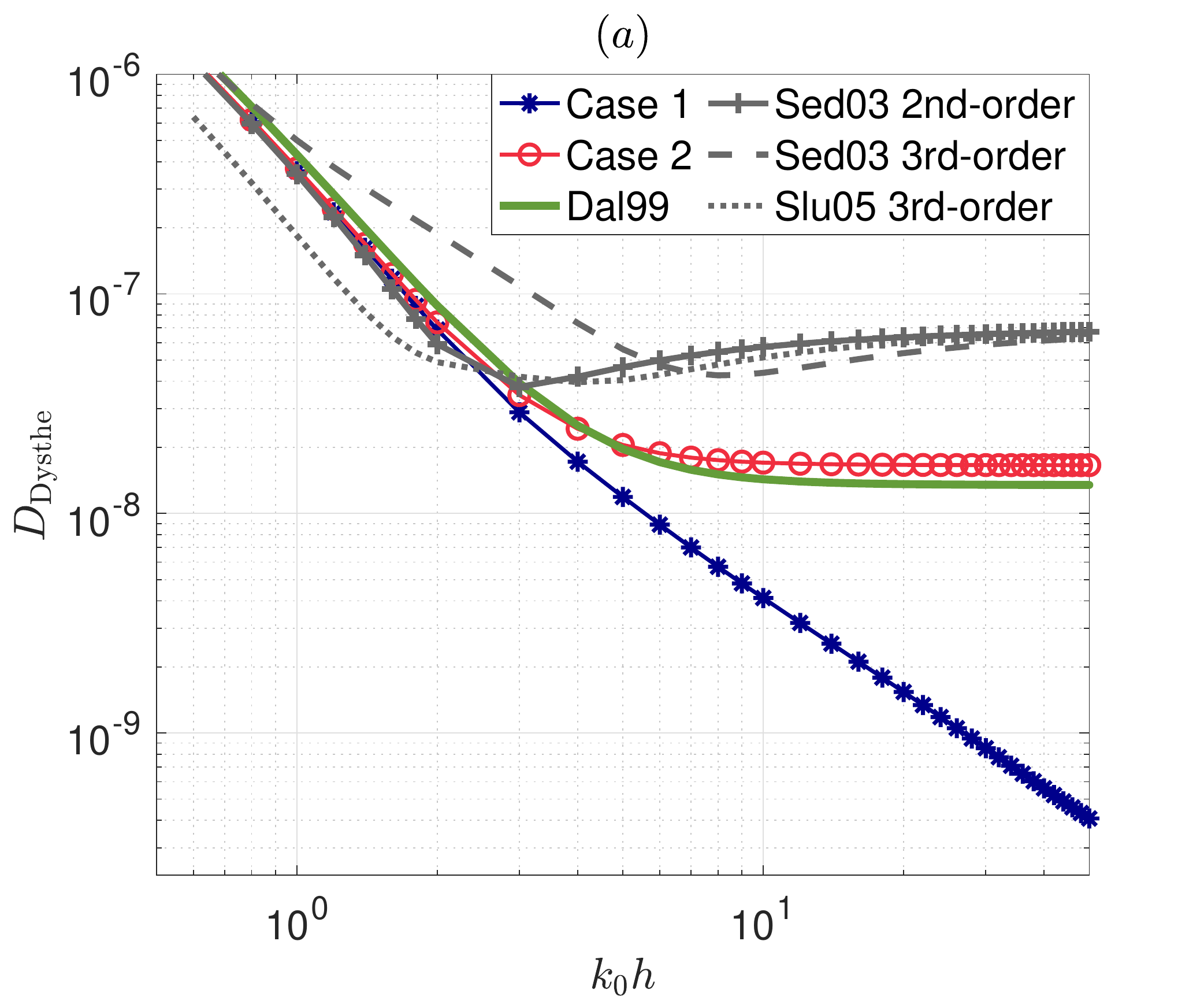}
\includegraphics[width=0.49\linewidth]{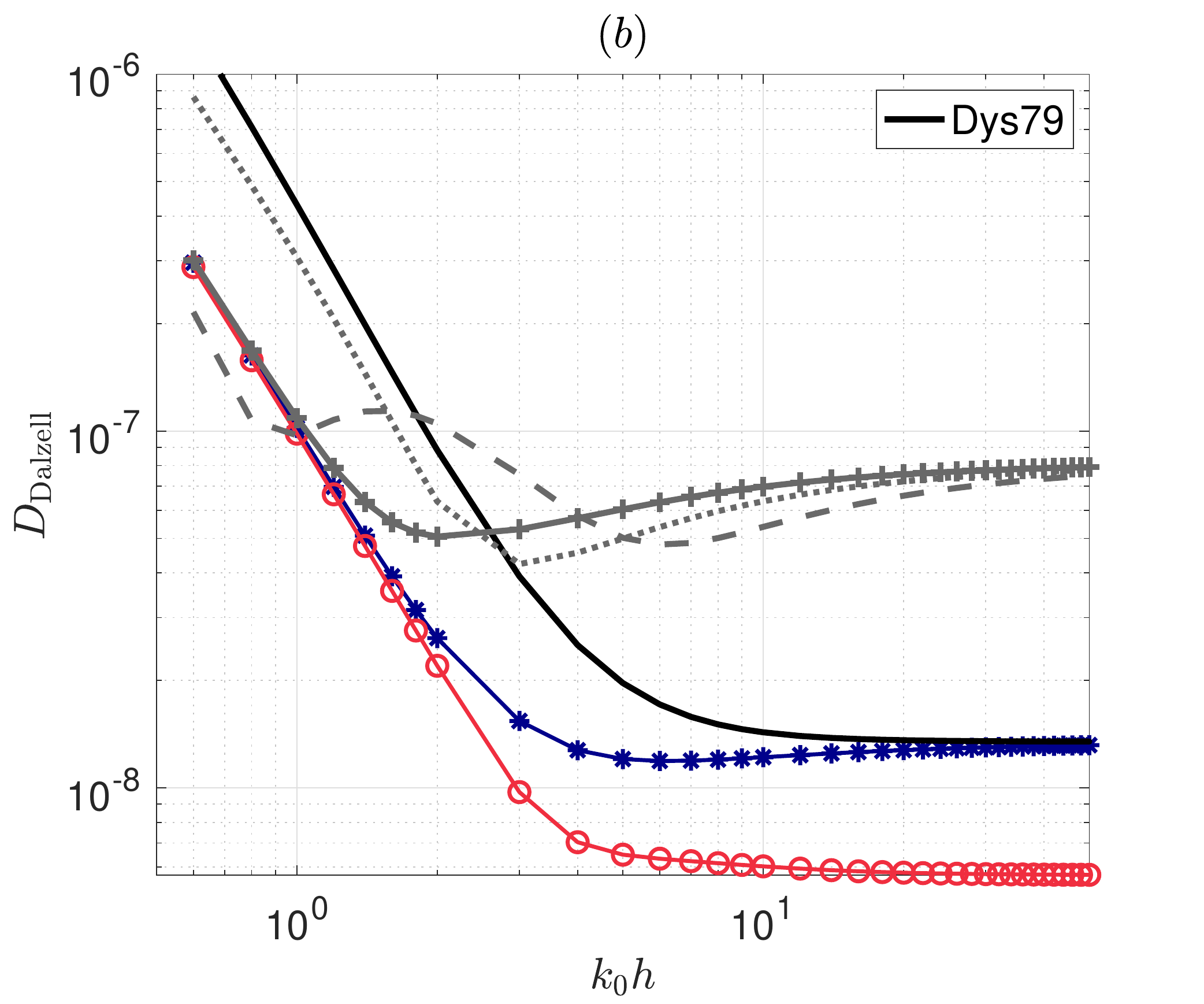}
    \caption{Deviation operator
with respect to the Dysthe expression ({\it a}) and to the Dalzell one  ({\it b}) for the derivative in time $\phi_{0t}$, as defined in the main text, for the different cases listed in Table~\ref{tab:table2}.} 
\label{fig:kh_timelike}
\end{figure*}

The results for $N= 30$ waves focusing at $t=0$ are shown in Fig.~\ref{fig:focusTL}. 
We see that the second-order expression in Sed03 reproduces well the waveform only for $k_0 h \le 1$, showing discrepancies with respect to Dalzell's solution at $k_0 h = 2$ and not converging to Dysthe's solution in deep-water.  
As for the space-like form, the expressions for $\phi_{0t}$ given by {\bf Case 1} [Eq.~(\ref{eq:IWtime_case1})] and {\bf Case 2} [Eq.~(\ref{eq:IWtime_case2})] have a similar behavior at all depths, providing in general approximations that are as accurate as the Dalzell solution in intermediate waters, and as the Dysthe expression at deep waters. As before, the third-order expressions provided by Eqs.~(\ref{eq:mean4orderTime}) and (\ref{eq:mean4orderSLUNtime}) are different and do not correspond to the other models.   
 Fig.~\ref{fig:kh_timelike} shows the deviation operator with respect to  the Dysthe expression, defined similarly to the spatial counterpart (see Sec.~\ref{sec:comparisons}) as $D_{\rm{Dysthe}} = N^{-1} \int (\phi_{0t}-\phi_{0t}^{\rm{Dysthe}})^2 \,dt$, and to the Dalzell one, $D_{\rm{Dalzell}} = N^{-1} \int (\phi_{0t}-\phi_{0t}^{\rm{Dalzell}})^2\, dt$ as a function of the non-dimensional water depth $k_0 h$. The integral is now calculated over $T = 80~T_0$, and the normalization coefficient is $N = (\omega_0 \lambda_{0DW})^{4}T$, with $\lambda_{0DW}$ calculated at deep water. 
As for the space-like case, it can be seen that the expression given by {\bf Case 2} [Eq.~\ref{eq:IWtime_case2})] is accurate at second-order at all depths as the second-order Dalzell solution, while the one given by {\bf Case 1} [Eq.~(\ref{eq:IWtime_case1})] is the only expression that converges to the Dysthe term in the deep-water limit, and is equivalent to {\bf Case 2} for $k_0 h<3$.

\section{Conclusion}
\label{conclusion}

During the evolution of surface gravity waves, fluid particles experience Stokes drift along the propagation direction of the waves, and a return flow in vertical and horizontal directions, which closes the water-mass transport. Both processes give rise to a mean flow at the surface that needs to be accounted for in order to accurately predict the transport of pollutants~\cite{dibenedetto2020,cozar2014,christensen2018} (such as oil, microplastics, etc) and, more in general, the impact of waves evolution at small scale on the ocean circulation at large scale~\cite{babanin2006,Onink2019,vanSebille2020,Cunnigham2022}.
We have derived nonlocal (viz.~non-instantaneous) expressions of the mean flow [Eqs.~(\ref{eq:IWspace}) and (\ref{eq:IWtime_case1})] that correctly converge to the deep-water limit at third-order in steepness, while being equivalent to second-order formulations in intermediate waters. 
We have included these expressions in an envelope evolution equation at fourth-order in steepness in both, space-like [Eq.~(\ref{eq:HONLSspace2})] and time-like [Eq.~(\ref{eq:HONLStimeFINAL})] formulations. 
We emphasize that the time-like form is relevant to study the evolution of unidirectional wave groups in space, thus, for modelling experiments in water wave flumes at high accuracy in arbitrary depths. Future work will be focusing on the experimental validation of our results in different water depth regimes.  

\begin{acknowledgments}
A.~G., J.~K. and M.~B. acknowledge financial support from the Swiss National Science Foundation (Project No. 200020-175697). D.E. acknowledges financial support from the Swiss National Science Foundation (Fellowship P2GEP2-191480). A. C. acknowledges
support from Kyoto University’s Hakubi Center for
Advanced Research.
\end{acknowledgments}

\section*{Author declarations}

\subsection*{Conflict of Interest Statement} 

The authors have no conflicts to disclose.

\subsection*{Author Contributions}

{\bf Alexis Gomel}: conceptualization (equal); formal analysis (equal); investigation (equal); writing/original draft (supporting); writing/review and editing (equal). 
{\bf Corentin Montessuit}: formal analysis (equal); investigation (equal). 
{\bf Andrea Armaroli:} conceptualization (equal); formal analysis (equal); investigation (equal);
writing/review and editing (equal).
{\bf Debbie Eeltink}: formal analysis (equal); writing/review and editing (equal). 
{\bf Amin Chabchoub}: supervision (equal); writing/review and editing (equal).
{\bf J\'er\^ome Kasparian}: supervision (equal); funding acquisition (lead); writing/review and editing (equal).
{\bf Maura Brunetti}: conceptualization (equal); formal analysis (equal); investigation (equal); methodology (lead); supervision (equal); writing/original draft (lead); writing/review and editing (equal).

\section*{Data Availability Statement}

The data that support the findings of
this study and the matlab scripts are available from the corresponding author upon reasonable
request.

\begin{widetext}
\appendix
\section{Dispersion and nonlinear coefficients} 
\label{app:A}

The dispersion coefficients are:
\begin{eqnarray}
\hat\alpha &=& \frac{1}{2}\omega''(k_0)  = \alpha \, c_\mathrm{g}^3\\ 
\alpha &=& -\frac{1}{2}k''(\omega_0) = -\frac{1}{2\omega_0 c_g}\left[1-\frac{g h}{c^2_g}(1-\kappa\sigma)(1-\sigma^2)\right] \\
\hat\alpha_3 &=& \frac{1}{6}\omega'''(k_0) \\
&=&\frac{\omega_0}{48 k_0^3\sigma} \left[3\sigma+\kappa(1-\sigma^2)\left(-3+\kappa\left(-\frac{3}{\sigma}+
\frac{3\kappa}{\sigma^2}+13\kappa-15\kappa(1-\sigma^2)-9\sigma\right)\right)\right]\nonumber \\
\alpha_3 &=& -\frac{1}{6}k'''(\omega_0) =  \frac{\hat\alpha_3}{c_\mathrm{g}^4}-2\alpha^2 c_\mathrm{g}\\
k'''(\omega_0) &=& -\frac{1}{c_\mathrm{g}^4}\pdern{c_\mathrm{g}}{k}{2} +
\frac{3}{c_\mathrm{g}^5}\left(\pder{c_\mathrm{g}}{k}\right)^2 \\
\hat\alpha_4 &\equiv& \frac{1}{24}\pdern{\omega}{k}{4} 
= \frac{\omega_0}{384 k_0^4}[-15 + 12\kappa^2 + 46\kappa^4 \\
&&+4\kappa(3 -5\kappa^2)\coth\kappa+2\kappa^2(3 + 10\kappa^2) \coth^2\kappa
+12\kappa^3 \coth^3\kappa -15\kappa^4 \coth^4\kappa\nonumber \\
&& +\kappa \sigma (-12+68\kappa^2 +3\kappa \sigma (-6-52\kappa^2+5\kappa\sigma
(-4+7\kappa\sigma)))] \nonumber \\
\alpha_4 &\equiv & \frac{1}{24}\pdern{k}{\omega}{4} = 
-\frac{1}{24 c_\mathrm{g}^5} \pdern{c_\mathrm{g}}{k}{3} +
\frac{5}{12 c_\mathrm{g}^6} \pdern{c_\mathrm{g}}{k}{2}\pder{c_\mathrm{g}}{k} - 
\frac{5}{8c_\mathrm{g}^7} \left(\pder{c_\mathrm{g}}{k}\right)^3 \nonumber \\
&=&-\frac{\hat\alpha_4}{c_\mathrm{g}^5} +
\frac{5}{c_g^6} \hat\alpha_3\hat\alpha - 
\frac{5}{c_\mathrm{g}^7} \hat\alpha^3
\end{eqnarray}
where $\kappa = k_0 h$, $\sigma = \tanh\kappa$ and $c_\mathrm{g}$ is the group velocity: 
\begin{equation}
c_g \equiv \pder{\omega}{k} = \frac{g}{2\omega_0}[\sigma+\kappa(1-\sigma^2)]
\end{equation}

The third-order nonlinear coefficients are:
\begin{eqnarray}
\hat\beta_D &=& -\frac{\omega_0 k_0^2}{16\sigma^4}(2\sigma^6 -13\sigma^4+12\sigma^2-9) \label{eq:hatbetaD}\\
\hat\beta &=& \beta\, {c_\mathrm{g}}= \hat\beta_D  +\omega_0 k_0^2 \frac{\mu_g^2}{8\sigma^2\nu}= \hat\beta_D  
-  \frac{k_0\mu_g}{4\sigma h} D \\
\beta_D &=& \frac{\hat\beta_D}{c_g}\\
\beta &=& \frac{\omega_0 k_0^2}{16\sigma^4 c_g}\left\{9-10\sigma^2+9\sigma^4-\frac{2\sigma^2c^2_g}{gh-c_g^2}\left[4\frac{c^2_p}{c^2_g}+4\frac{c_p}{c_g}(1-\sigma^2)+\frac{g h}{c^2_g}(1-\sigma^2)^2\right]\right\}
\end{eqnarray}
where, interestingly, the coefficients in the curly brackets can be expressed in terms of $\kappa$: $c_g/c_p = (\sigma +\kappa(1-\sigma^2))/(2\sigma)$, $gh/c_g^2 = (c_p^2/c_g^2)\kappa/\sigma$,  $2\sigma^2c^2_g/(gh-c_g^2)= 
2\sigma (c_g^2/c_p^2)/(\kappa/\sigma-c_g^2/c_p^2)$ and 
\begin{eqnarray}
D &=& -h \frac{\omega_0}{2} \frac{k_0 \mu_g}{\sigma \nu} = 
\frac{\omega_0}{2} \frac{\kappa}{2\sigma}\frac{2+(1-\sigma^2)c_g/c_p}{\kappa -\sigma c_g^2/c_p^2} 
= \frac{D'}{1-c_g^2/(gh)}
\label{eq:Ddef} \\
D' &=& \frac{\omega_0}{2\sigma} (1+C_{FD}) \\
\mu_g &=& \frac{2\sigma}{\omega_0}(2\omega -kc_g(\sigma^2-1)) = (\sigma^2 -1)^2 \kappa - \sigma(\sigma^2-5) = 4\sigma (1+C_{FD})\label{eq:mug}\\     
\nu &=& \frac{4k_0\sigma}{g}(c_g^2-gh)= [(\sigma+1)^2\kappa-\sigma][(\sigma-1)^2\kappa-\sigma] \label{eq:nu}\\
C_{FD} &=& \frac{\omega_0 c_g}{g\sinh(2\kappa)} \label{eq:CFD}
\end{eqnarray}

The higher-order nonlinear coefficients are:
\begin{eqnarray}
\beta_{21} &=& \frac{\hat\beta_{21}}{c_\mathrm{g}^2}- 4\alpha\beta c_\mathrm{g}\\
\beta_{22} &=& \frac{\hat\beta_{22}}{c_\mathrm{g}^2}- 2\alpha\beta c_\mathrm{g}\\
\hat\beta_{21} &=& \omega_0 k_0 Q_{41S} \\
\hat\beta_{22} &=& \omega_0 k_0 Q_{42S} \\
\mathcal{B}_{21} &=& \omega_0 k_0 \tilde Q_{41}/c_g^2 - 4\alpha\beta_D c_g\\
\mathcal{B}_{22} &=& \omega_0 k_0 \tilde Q_{42}/c_\mathrm{g}^2 - 2\alpha\beta_D c_\mathrm{g}
\end{eqnarray}
where $Q_{41S}$, $Q_{42S}$, $\tilde Q_{41}$, $\tilde Q_{42}$ are given in App.~\ref{app:B}.

\section{Notation used in Sedletsky (2003) (Sed03)} 
\label{app:B}

The main coefficients in Sedletsky's notation are~\cite{Sedletsky2003}:
\begin{eqnarray}
Q_{41} &=& \tilde Q_{41}-\frac{\mu_g}{\nu}\tilde q_{40} \label{sed1true}\\
Q_{42} &=& \tilde Q_{42}+\frac{\mu_g}{\nu}\tilde q_{40} \label{sed2true}\\
\tilde q_{40} &=& \frac{1}{32\sigma^3\nu}[ (\sigma^2-1)^5 \kappa^4 -4\sigma(2\sigma^4+9\sigma^2+5)(\sigma^2-1)^2\kappa^3 \nonumber \\
&&+2\sigma^2(9\sigma^4+16\sigma^2-9)(\sigma^2-1)\kappa^2 \nonumber \\
&&-4\sigma^3(4\sigma^4-9\sigma^2-7)\kappa + 5\sigma^4(\sigma^2-5)] \label{eq:tildeq40}\\
\tilde Q_{41}&=& \tilde q_{41}  \nonumber \\ 
&=& \frac{1}{16\sigma^5\nu}
[ (2\sigma^6-11\sigma^4-10\sigma^2+27)(\sigma^2-1)^3 \kappa^3 -\sigma(6\sigma^8-21\sigma^6+9\sigma^4-43\sigma^2+81)(\sigma^2-1)\kappa^2 \nonumber \\
&&+\sigma^2(6\sigma^8-15\sigma^6-77\sigma^4+71\sigma^2-81)\kappa -\sigma^3(\sigma^2+1) (2\sigma^4-7\sigma^2-27)] 
\label{eq:tildeQ41}\\
\tilde q_{42} &=& \frac{1}{32\sigma^5\nu}
[ (4\sigma^6-13\sigma^4+10\sigma^2-9)(\sigma^2-1)^3 \kappa^3 -\sigma(12\sigma^8-51\sigma^6+17\sigma^4-\sigma^2-9)(\sigma^2-1)\kappa^2 \nonumber \\
&&+\sigma^2(12\sigma^8-67\sigma^6+33\sigma^4-\sigma^2-9)\kappa -\sigma^3 (4\sigma^6-29\sigma^4+42\sigma^2-9)] \\
q_3 &=& - \frac{\hat\beta}{\omega_0 k_0^2}\\
\tilde Q_{42} &=& \tilde q_{42} -2 \frac{c_g}{c_p} q_3 \label{eq:tildeQ42}
\end{eqnarray} 
The final expressions for $Q_{41}$ and $Q_{42}$ are given in Eqs.~(67) and (68) in Ref.~\onlinecite{Sedletsky2003} and reported here for completeness: 
\begin{eqnarray}
\mathcal{Q}_{41} &=& \frac{1}{32\sigma^5\nu^2}\{(3\sigma^6-20\sigma^4-21\sigma^2+54)(\sigma^2-1)^5 \kappa^5  \nonumber \\ 
&& -\sigma(11\sigma^8-99\sigma^6-61\sigma^4+7\sigma^2+270)(\sigma^2-1)^3 \kappa^4 \nonumber \\ &&+2\sigma^2(\sigma^2-1)(7\sigma^{10} -58\sigma^8+38\sigma^6+52\sigma^4-181\sigma^2+270)\kappa^3  \nonumber \\
&&- 2\sigma^3(3\sigma^{10} +18\sigma^8-146\sigma^6 -172\sigma^4 +183\sigma^2 -270)\kappa^2 \nonumber \\
&&-\sigma^4(\sigma^8-109\sigma^6+517\sigma^4+217\sigma^2+270)\kappa \nonumber \\ 
&&+ \sigma^5(\sigma^6-40\sigma^4+193\sigma^2+54)\} 
\label{Sedl1}\\
\mathcal{Q}_{42} &=& \frac{1}{32\sigma^5\nu^2}\{-(3\sigma^6+7\sigma^4-11\sigma^2+9)(\sigma^2-1)^5 \kappa^5  \nonumber \\ 
&&+ \sigma(11\sigma^8-48\sigma^6+66\sigma^4+8\sigma^2+27)(\sigma^2-1)^3 \kappa^4 \nonumber \\ &&-2\sigma^2(\sigma^2-1)(7\sigma^{10} -79\sigma^8+282\sigma^6-154\sigma^4-\sigma^2+9)\kappa^3  \nonumber \\
&&+ 2\sigma^3(3\sigma^{10} -63\sigma^8+314\sigma^6 -218\sigma^4 +19\sigma^2 +9)\kappa^2  \nonumber \\
&&+\sigma^4(\sigma^8+20\sigma^6-158\sigma^4-28\sigma^2-27)\kappa - \sigma^5(\sigma^6-7\sigma^4+7\sigma^2-9)\} 
\label{Sedl2}
\end{eqnarray}

We have verified that they are equivalent to the expressions obtained using Eqs.~(\ref{sed1true})-(\ref{sed2true}).  

From Eqs.~(\ref{Sedl1})-(\ref{Sedl2}), 
in the deep-water limit $\kappa \to \infty$, $\sigma \to 1$, $\nu \to 1-4\kappa$, $\mu_g \to 4$, $\mathcal{Q}_{41} \to 768/(32\cdot 16) = 3/2$ and  $\mathcal{Q}_{42} \to 128/(32\cdot 16) = 1/4$ recovering the Dysthe result for such terms. 

As suggested in Ref.~\onlinecite{gandzha2014}, the above expressions can be modified to agree with the results presented in Ref.~\onlinecite{Slunyaev2005}. 
However, we have verified that the (small) modification suggested in Ref.~\onlinecite{gandzha2014} missed a factor 2, the right one being the following: 
\begin{eqnarray}
\tilde q_{40S} &=& \tilde q_{40} + \frac{\Delta}{2} \frac{\nu}{\mu_g} \label{eq:q40S}\\
Q_{41S}&=& Q_{41} -\frac{\Delta}{2} \label{eq:beta21}\\
Q_{42S}&=& Q_{42} +\frac{\Delta}{2} 
\label{eq:beta22}
\end{eqnarray}
where the term $\Delta$ is defined in Ref.~\onlinecite{gandzha2014}:
\begin{eqnarray}
\Delta &=& -\frac{\sigma^2-1}{16\sigma^3\nu}[(\sigma^2-1)^3(3\sigma^2+1)\kappa^3 -\sigma(\sigma^2-1)(5\sigma^4-18\sigma^2-3)\kappa^2 \nonumber \\
&+&\sigma^2(\sigma^2-1)(\sigma^2-9)\kappa+\sigma^3(\sigma^2-5)]
\end{eqnarray}
In the deep-water limit $\kappa \to \infty$,  $Q_{41S} \to  3/2$ and  $Q_{42S} \to 1/4$, since $\Delta \to 0$, thus recovering the Dysthe result for such terms.

Note, however, that the final equations that include the mean flow (namely, Eqs. (\ref{eq:HONLSspace2}) and (\ref{eq:HONLStimeFINAL})) are not affected by such ambiguity, since their high-order nonlinear coefficients only depend on $\tilde Q_{41}$ and $\tilde Q_{42}$.  

\section{Notation used in Slunyaev (2005) (Slu05)} 
\label{app:C}

One can use the notation in Ref.~\onlinecite{Slunyaev2005}, where $\hat\beta_{21} = \omega_0 k_0 Q_{41S}$ and $\hat\beta_{22} = \omega_0 k_0 Q_{42S}$, and $Q_{41S} = (h^3 \omega_0) \tilde \alpha_{21}/(\kappa^3 \sigma^2)$, 
$Q_{42S} = (h^3 \omega_0) \tilde \alpha_{22}/(\kappa^3 \sigma^2)$, 
with  $\tilde \alpha_{21} = \tilde \rho_{21} -\tilde\rho_{12}\gamma_2$ and  $\tilde \alpha_{22} = \tilde\rho_{22}+\tilde\rho_{12}\gamma_2$, $\tilde\rho_{21}= P_{21}+s\beta_1\gamma_1$, 
$\tilde\rho_{22}= P_{22}-s\beta_1\gamma_1$. 
Here $\beta_1 = -\hat\alpha$ and the other coefficients are given by\footnote{We have expressed Slunyaev's coefficients in terms of Sedletsky's expressions defined in App.~\ref{app:A} and \ref{app:B} whenever possible.}
\begin{eqnarray}
h^3 \omega_0 P_{21} &=& \left(\kappa^2 \frac{(\sigma^2-1)(-4\sigma^4+3\sigma^2+1)}{8\sigma^2} + 
\kappa \frac{4\sigma^4-9\sigma^2+3}{4\sigma} +\frac{-4\sigma^2+19}{8}\right) h^3 \omega_0 \gamma_1  \nonumber\\
&&+ \kappa^2 \frac{-\sigma^4 + 3}{2(\sigma^2+1)} h \omega_0 \chi_2 + 
\left( \kappa^2  \frac{-3\sigma^6+7\sigma^4-9\sigma^2-3}{4\sigma(\sigma^2+1)} + 
3\kappa \frac{\sigma^4-5}{4(\sigma^2+1)}\right) h^2 \omega_0 \chi_1  \nonumber \\
&& + \kappa^4 
\frac{(\sigma^2-1)(11\sigma^4-12\sigma^2-3)}{16\sigma} +\kappa^3\frac{-11\sigma^4 + 40\sigma^2 - 9}{16}\\ 
h^3 \omega_0 P_{22} &=& \left(-\kappa^2 \frac{(\sigma^2-1)^2}{8} + 
\kappa \frac{\sigma^4-5\sigma^2+2}{4\sigma} +\frac{-\sigma^2+8}{8}\right) h^3 \omega_0 \gamma_1  \nonumber\\
&& + 
\left( \kappa^2  \frac{(\sigma^2-1)(\sigma^4+3)}{4\sigma(\sigma^2+1)} - 
\kappa \frac{\sigma^4+3}{4(\sigma^2+1)}\right) h^2 \omega_0 \chi_1 \nonumber\\
&& + \kappa^4  \frac{(\sigma^2-1)(-3\sigma^4-8\sigma^2+3)}{32\sigma} + 
3\kappa^3 \frac{\sigma^4 -1}{32} \\
h^2 \omega_0 s&=& \kappa^2 \frac{\sigma^2-1}{2} \\
V_d^2 &=& gh - c_g^2 = -\frac{g\nu}{4k_0\sigma}\\
h^3 \omega_0 \gamma_1 &=& h^3 \omega_0 \frac{k_0^2 c_g (\sigma^2-1)-2\omega_0 k_0}{4V_d^2} = \frac{\kappa^3\sigma\mu_g}{2\nu}\\
\gamma_2 &=& \frac{1}{V_d^2}\left[2 c_g \gamma_1 \beta_1 +k_0^2 \beta_1 \frac{(\sigma^2-1)}{4} + \frac{\omega^2 -k_0^2c_g^2(\sigma^2-1)}{4\omega_0}\right] \\
\tilde\rho_{12} &=& \frac{2\omega_0 k_0 -k_0^2 c_g(\sigma^2-1)}{2\omega_0} = \frac{k\mu_g}{4\sigma}\\
h^2 \omega_0 \chi_1 &=& 3 \kappa^2 \frac{\sigma^4-1}{8 \sigma^2} \\
h\omega_0 \chi_2 &=& \left(\kappa  \frac{-\sigma^3+3}{\sigma} +1\right) \frac{h^2\omega_0\chi_1}{\kappa} + 
\frac{3\kappa^2 (\sigma^2-1)(3\sigma^2+1)}{16\sigma} + 9\kappa\frac{1-\sigma^4}{16\sigma^2}
\end{eqnarray}
Using the same notation as in Sed03 for the reconstruction at leading order  of the surface elevation, $\eta(\chi,t) = \frac{1}{2} [U(x,t) \exp(i (k_0 \chi- \omega_0 t)) + {\rm{c.~c.}}]$, the mean flow term is written as (see Eqs.~(32), (43), (45) in Ref.~\onlinecite{Slunyaev2005}):
\begin{equation}
     \pder{\phi_0}{x} =   \frac{\omega_0}{2} \frac{k_0 \mu_g}{\sigma \nu}  |U|^2 - i \frac{c_p^2}{\sigma^2} \left(\gamma_2+\frac{\beta_1\gamma_1}{c_g}\right) \left(U \pder{U^*}{x}- U^* \pder{U}{x}\right)
\label{eq:mean4orderSLUN}
\end{equation}
Note that, while the first term is equal to the corresponding term in Eq.~(\ref{eq:mean4order}) from Sedletsky's derivation, the second one has a different coefficient, since $ {4\omega_0 \sigma}\tilde q_{40}/\nu  \not =  c_p^2{\sigma^2} (\gamma_2+\beta_1\gamma_1/c_g)/{\sigma^2}$ (even if we take into account the correction $\tilde q_{40S}$ in Eq.~(\ref{eq:q40S})). 

Using Eq.~(\ref{eq:cgphi0}), the derivative in time of $\phi_0$ is 
\begin{equation}
     \pder{\phi_0}{t} =  -c_g \frac{\omega_0}{2} \frac{k_0 \mu_g}{\sigma \nu}  |U|^2 - i \frac{c_p^2}{\sigma^2} \left(\gamma_2+\frac{\beta_1\gamma_1}{c_g}\right) \left(U \pder{U^*}{t}- U^* \pder{U}{t}\right)
\label{eq:mean4orderSLUNtime}
\end{equation}

\section{Example of sea state realization}
\label{app:D}

Following the definitions in Ref.~\onlinecite{LiLi2021}, we consider a Gaussian amplitude spectrum, denoted by $S(k)$, given by
\begin{equation}
S(k) = \exp\left(-\frac{(k - k_0)^2}{2k_w^2}\right) 
\end{equation}
for $k>0$, where $k_0 =  0.0277$~m$^{-1}$ and $k_w=0.27~k_0$ is the dimensional bandwidth. Other spectra, like JONSWAP or Pierson-Moskowitz, can also be used. This yields the following surface elevation at first order in steepness: 
\begin{equation}
\eta_1(x,t) = \frac{A_p}{2} \frac{\int_0^\infty S(k) e^{i[k(x-x_f)-\omega (t-t_f)])} dk} {\int_0^\infty S(k) dk}   + c.c. 
\end{equation}
where $A_p$, $x_f$, $t_f$ are the amplitude, position and time for the group at linear focus, with steepness given by $\varepsilon = A_p k_0 = 0.3$. An example of sea state realization at second order in steepness, obtained using the Dalzell development~\cite{dalzell1999}, is shown in Fig.~\ref{fig:focusApp} for the case of a focused wave group at $x_f=0$, $t_f = 0$ for $k_0 h = 1.5$. The power spectrum can be obtained as~\cite{young2020} $P(k) = |\hat\eta|^2/(2 dk)$, where $\hat\eta$ is the Fourier transform in space of the surface elevation, from which the significant wave height $H_s$ can be obtained as $H_s = 4\sqrt{m_0} = 5.2$~m, $m_0$ being equal to the area under the power spectrum curve.

\begin{figure*}[!]
\includegraphics[width=0.495\textwidth]{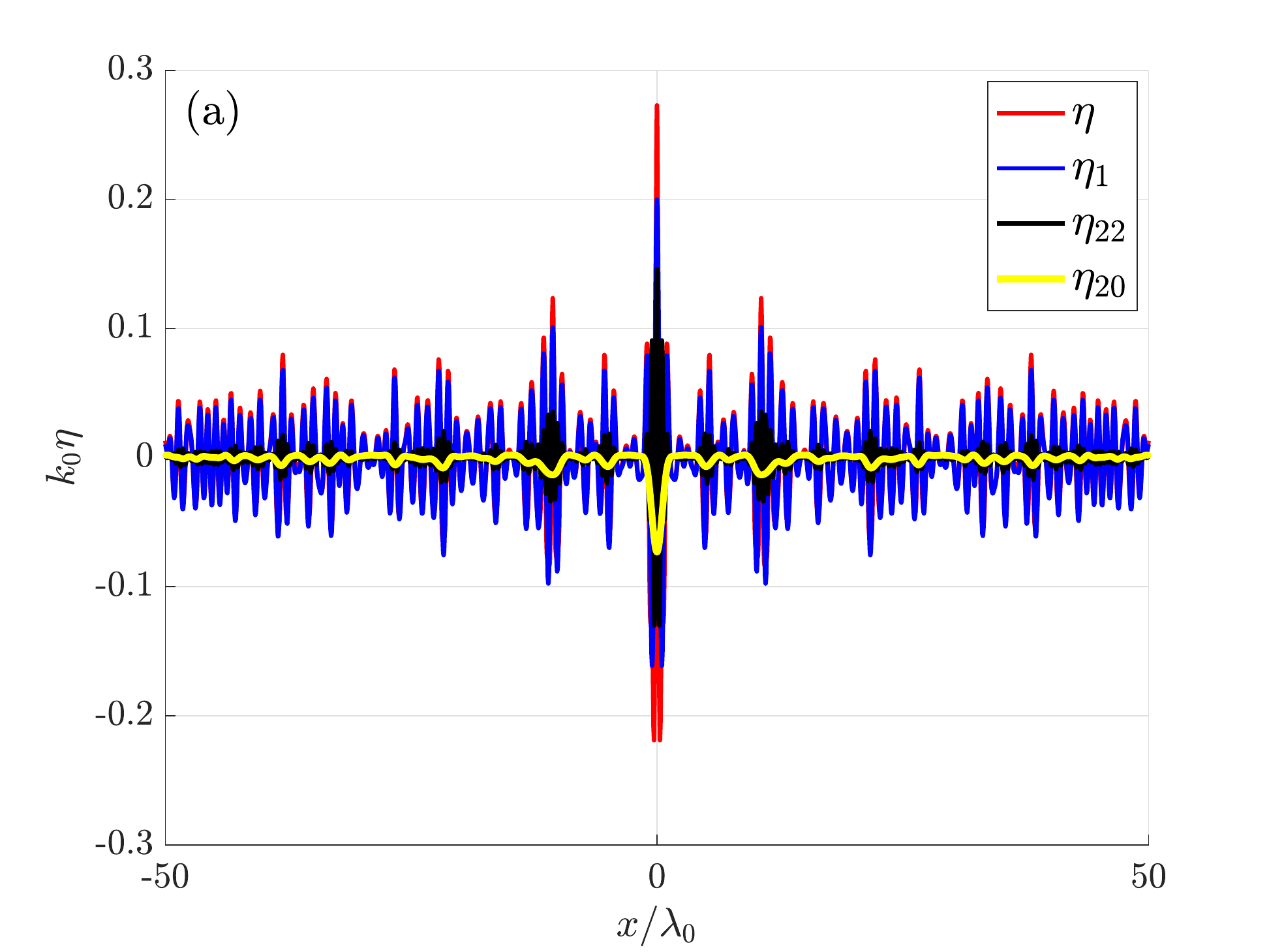}
\includegraphics[width=0.495\textwidth]{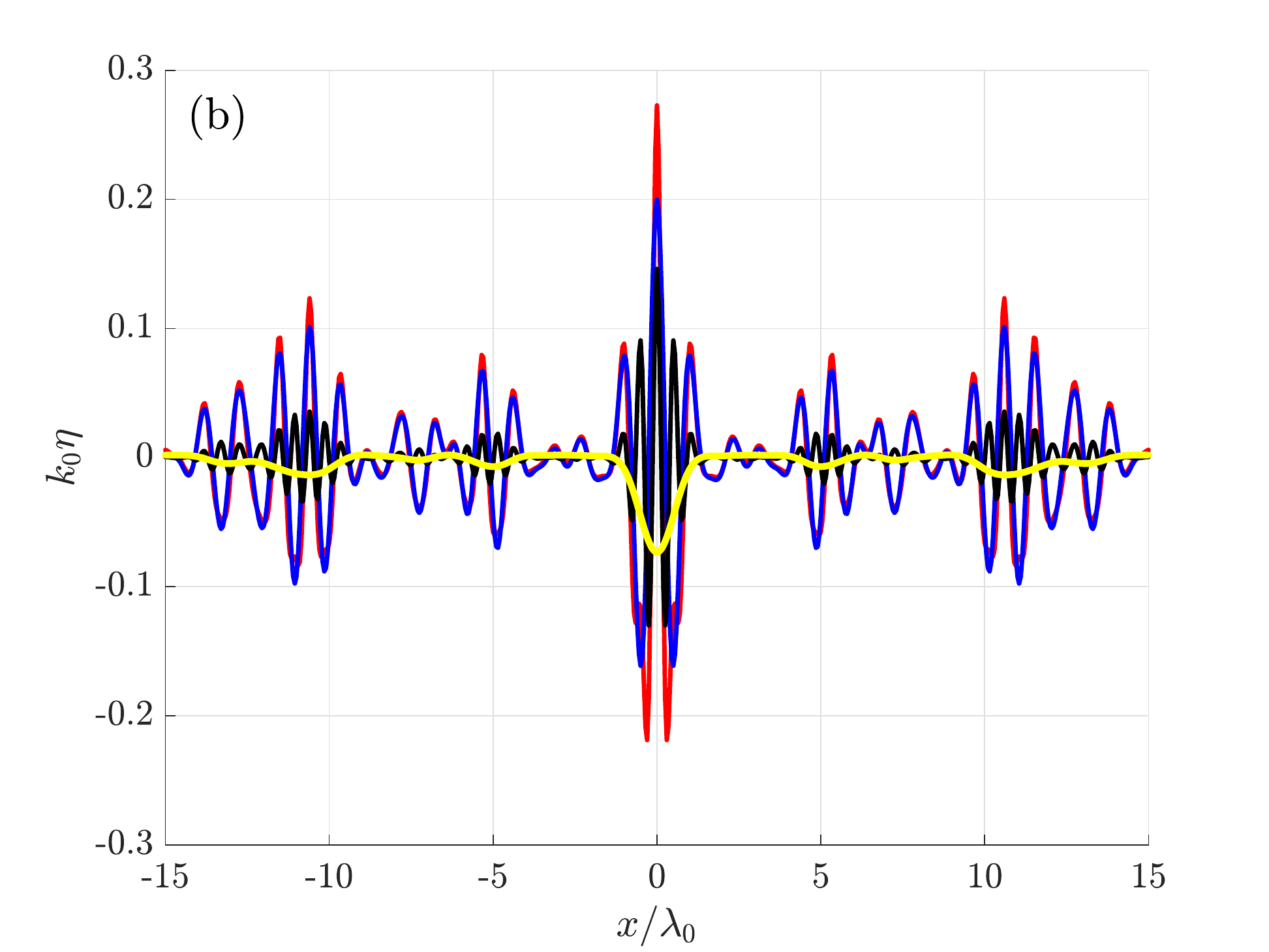}
    \caption{Sea surface elevation $\eta$ and its components at first ($\eta_1$) and second order in steepness ($\eta_{22}$ and the set-down $\eta_{20}$)  in the case of a focused wave group at the origin for $k_0 h= 1.5$: ({\it a}) full range of considered spatial values; ({\it b}) same as ({\it a}), zoomed on $30 \lambda_0$.}
\label{fig:focusApp}
\end{figure*}

\end{widetext}

%

%
%

%


\bibliography{waves}
\end{document}